\documentclass[sigconf,screen,nonacm]{acmart}
\usepackage{color, colortbl, xcolor}
\usepackage{url}
\usepackage{subcaption}
\usepackage{textcomp}
\usepackage{soul}
\usepackage{multirow}
\usepackage{enumitem}
\usepackage{mathtools}
\usepackage{siunitx}
\usepackage{xeCJK}

\usepackage{booktabs} 

\usepackage{array}
\usepackage{xcolor}

\newcommand*{\rowstyle}[1]{
  \gdef\@rowstyle{#1}%
  \@rowstyle\ignorespaces%
}

\newcolumntype{=}{
  >{\gdef\@rowstyle{}}%
}

\newcolumntype{+}{
  >{\@rowstyle}%
}

\usepackage{arydshln}
\definecolor{linkColor}{RGB}{6,125,233}
\definecolor{green}{rgb}{0.0, 0.65, 0.31}
\definecolor{bleudefrance}{rgb}{0.19, 0.55, 0.91}
\definecolor{ceruleanblue}{rgb}{0.16, 0.32, 0.75}
\definecolor{grey}{HTML}{969696}
\definecolor{violet}{HTML}{756bb1}
\definecolor{dgrey}{HTML}{01665e}
\definecolor{lgrey}{HTML}{5ab4ac}
\definecolor{dgreen}{HTML}{005a32}
\definecolor{purple}{HTML}{ae017e}

\definecolor{editCol}{HTML}{0000FF}
\definecolor{maskCol}{HTML}{c51b7d}
\definecolor{lrColor}{HTML}{8856a7}
\definecolor{trColor}{HTML}{d01c8b}
\definecolor{ctColor}{HTML}{4dac26}
\definecolor{brickred}{HTML}{f03b20}
\definecolor{improveCol}{HTML}{253494}
\definecolor{worsenCol}{HTML}{d7191c}
\definecolor{DarkBlue}{HTML}{00008B}
\definecolor{mscolor}{HTML}{01665e}
\definecolor{nmscolor}{HTML}{bf812d}
\definecolor{lgreen}{HTML}{ccece6}
\definecolor{dolive}{HTML}{308014}

\colorlet{tablerowcolor4}{gray!50} 

\newcommand*{\textlabel}[2]{%
  \edef\@currentlabel{#1}
  \phantomsection
  #1\label{#2}
}

\newcommand{\ra}[1]{\renewcommand{\arraystretch}{#1}}
\colorlet{tableheadcolor}{gray!25} 
\colorlet{tablerowcolor}{gray!10} 
\colorlet{tablerowcolor2}{gray!45} 
\colorlet{tablerowcolor3}{gray!25} 

\newcolumntype{a}{>{\columncolor{tablerowcolor}}r}
\definecolor{aicolor}{HTML}{018571}
\definecolor{occolor}{HTML}{ff7799}

\definecolor{aicolor}{HTML}{fc8d62}
\definecolor{occolor}{HTML}{253494}

\newif{\ifhidecomments}
\hidecommentstrue
\ifhidecomments
    \newcommand{\yunhao}[1]{}
    \newcommand{\kejia}[1]{}
    \newcommand{\yuqi}[1]{}
    \newcommand{\yajing}[1]{}
    \newcommand{\renwen}[1]{}
    \newcommand{\talayeh}[1]{}
\else
    \newcommand{\yunhao}[1]{\textbf{\small\sffamily{\textcolor{DarkBlue}{[#1 -- Yunhao]}}}}
    \newcommand{\kejia}[1]{\textbf{\small\sffamily{\textcolor{dgreen}{[#1 -- Kejia]}}}}
    \newcommand{\yuqi}[1]{\textbf{\small\sffamily{\textcolor{purple}{[#1 -- Yuqi]}}}}
    \newcommand{\yajing}[1]{\textbf{\small\sffamily{\textcolor{ceruleanblue}{[#1 -- Yajing]}}}}
    \newcommand{\renwen}[1]{\textbf{\small\sffamily{\textcolor{bleudefrance}{[#1 -- Renwen]}}}}
    \newcommand{\talayeh}[1]{\textbf{\small\sffamily{\textcolor{violet}{[#1 -- Talayeh]}}}}
  \fi

\renewcommand{\textrightarrow}{$\rightarrow$}

\newcommand{\aic}{AICC}

\colorlet{tableheadcolor}{gray!25} 

\definecolor{neutralCol}{HTML}{dd1c77}
\definecolor{neutralGreen}{HTML}{31a354}
\definecolor{NewBlue}{HTML}{1879ba}
\definecolor{bleudefrance}{rgb}{0.19, 0.55, 0.91}  
\definecolor{AfTrColor}{HTML}{0868ac}  
\definecolor{BfTrColor}{HTML}{a8ddb5}  

\definecolor{AfCtColor}{HTML}{b10026}  
\definecolor{BfCtColor}{HTML}{fd8d3c}

\graphicspath{ {figures/} }

\newcommand{\para}[1]{\vspace{0.3em}\noindent\textbf{#1}~}

\newcommand{\provisionalparaphrase}[1]{%
  \begin{quote}
  \small\raggedright\textcolor{orange!55!black}{\textit{[PROVISIONAL PARAPHRASE---Translate before submission:] "#1"}}
  \end{quote}
}

\definecolor{qidxCol}{HTML}{4d4d4d}

\newcommand{\qidxstyle}[1]{{\normalfont\textcolor{qidxCol}{#1}}}

\newcommand{\qidx}[1]{\qidxstyle{(#1)}}

\newcommand{\iqo}[3]{``\textit{#1}'' ({#3})}

\renewcommand{\provisionalparaphrase}[2][]{%
  \begin{quote}
  \small\raggedright\textcolor{orange!55!black}{%
    \textit{[PROVISIONAL PARAPHRASE---Translate before submission:] "#2"}}%
  \if\relax\detokenize{#1}\relax\else\ \qidx{#1}\fi
  \end{quote}}

\usepackage{tcolorbox}
\tcbuselibrary{breakable,skins}

\usepackage{xeCJK}

\newtcolorbox{promptbox}[1][]{%
  breakable,
  enhanced jigsaw,
  colback=black!2,
  colframe=black!35,
  boxrule=0.4pt,
  left=6pt, right=6pt, top=6pt, bottom=6pt,
  before skip=6pt, after skip=6pt,
  fonttitle=\sffamily\bfseries\small,
  #1
}

\newcommand{\psec}[1]{%
  \par\addvspace{4pt}%
  {\sffamily\bfseries\small [#1]}\par\addvspace{2pt}}

\newcommand{\promptlists}{%
  \setlist[enumerate]{nosep,leftmargin=1.6em,labelsep=0.4em,topsep=2pt}%
  \setlist[itemize]{nosep,leftmargin=1.2em,labelsep=0.4em,topsep=2pt,label=\textendash}%
}

\AtBeginDocument{%
  \providecommand\BibTeX{{%
    \normalfont B\kern-0.5em{\scshape i\kern-0.25em b}\kern-0.8em\TeX}}}

\begin{document}

\title[When AI Companions Disappear]{When AI Companions Disappear}
\subtitle{Relational Continuity and Collective Contestation during China's National AI Regulatory Transition}
\author{Yunhao Yuan}
\orcid{0000-0002-1450-8572}
\affiliation{%
  \institution{Aalto University}
  \city{Espoo}
  \state{}
  \country{Finland}
}
\email{yunhao.yuan@aalto.fi}

\author{Kejia Zhang}
\orcid{0009-0008-8285-6043}
\affiliation{%
  \institution{The University of Edinburgh}
  \city{Edinburgh}
  \state{}
  \country{United Kingdom}
}
\email{k.zhang-61@sms.ed.ac.uk}

\author{Yuqi Niu}
\orcid{0009-0004-7624-4711}
\affiliation{%
  \institution{Shanghai Jiao Tong University}
  \city{Shanghai}
  \state{}
  \country{China}
}
\email{niuyuqi@sjtu.edu.cn}

\author{Yajing Wang}
\orcid{0000-0001-7823-5385}
\affiliation{%
  \institution{Aalto University}
  \city{Espoo}
  \state{}
  \country{Finland}
}
\email{yajing.wang@aalto.fi}

\author{Renwen Zhang}
\orcid{0000-0002-7636-9598}
\affiliation{
  \institution{Nanyang Technological University}
  \city{Singapore}
  \state{}
  \country{Singapore}}
\email{renwen.zhang@ntu.edu.sg}

\author{Talayeh Aledavood}
\orcid{0000-0002-0110-5694}
\affiliation{%
  \institution{Aalto University}
  \city{Espoo}
  \state{}
  \country{Finland}}
\email{talayeh.aledavood@aalto.fi}

\renewcommand{\shortauthors}{Yunhao Yuan et al.}

\begin{abstract}

AI model updates and service withdrawals can disrupt relationships with AI companions, but research has largely examined individual platform events. Less is known about users' responses when multiple providers implement shared national regulations. We examine users' reactions and collective contestation surrounding China's 2026 regulation of anthropomorphic AI interaction services. We collected RedNote discussions from April 10 to August 10, 2026, used a validated language model for relevance screening, and conducted qualitative content and thematic analyses of 89 posts, 1,425 comments, and 2,005 replies. Users retained, migrated, and reconstructed companions, finding that preserving conversation records did not necessarily restore shared memories or familiar interactions. They compared regulations, platform explanations, and implementations to assign responsibility. Solidarity emerged through mutual aid and appeals to respect other communities' attachments, while disputes over targets and tactics exposed contested terms of collective action. Infrastructural dependence connected relational continuity and collective contestation during this national regulatory transition.

\end{abstract}



\maketitle

\section{Introduction}
\yunhao{higlighted comments usage}
\kejia{}
\yuqi{}
\yajing{}
\renwen{}
\talayeh{}

\yunhao{\textbf{Word count (Max 12,000) (Abstract--Conclusion): \mainwordcount}}

\yunhao{\textbf{Word count (Max 150)(Abstract): 150}}

AI-powered companion chatbots (\aic{}s)
are conversational systems designed to offer personalized and emotionally responsive interactions \cite{de2026ai,ho2025potential,smith2025can,li2024finding}.
Through recurring interactions, intimate disclosures, and accumulated shared histories, \aic{} users may develop long-term relationships with companions they perceive as friends, partners, family members, or confidants~\cite{skjuve2021my,skjuve2022longitudinal,ta2020user,xie2023friend,ng2026love,li2024finding,zhang2026fragility}. 
Government rules that apply across \aic{}s can reshape these relationships, %
but \aic{} users encounter these requirements through the decisions of individual platforms. Providers control the accounts, models, memories, and interactional affordances through which companions remain accessible and recognizable. Across platforms, users therefore share a condition of infrastructural dependence: the continuation of a personally meaningful relationship relies on systems they do not fully control. When providers restrict access, alter relational features, or withdraw services, this dependence becomes consequential through the possible loss of a particular companion and shared history  \cite{lee2026large,banks2024deletion,de2024lessons}. Moving to another platform can change who controls these conditions, but the relationship remains subject to decisions beyond the user's control.

Prior research documents grief and perceptions of an altered or lost companion following feature removals and behavioral changes \cite{laestadius2024too,hanson2024replika,cagiltay2026rip}, as well as companion endings understood through deletion, departure, and death \cite{banks2024deletion}. In the Replika case, the removal of erotic role-play followed an enforcement action by the Italian Data Protection Authority, and subsequent research examined users' experiences of identity discontinuity and mourning \cite{de2024lessons,de2026mourning}. 
These studies largely center on disruptions within individual services \cite{de2024lessons,laestadius2024too,hanson2024replika,de2026mourning,lai2026please}.
China's 2026 transition involved multiple providers introducing different restrictions, withdrawals, and transition arrangements under a shared national framework. This variation raises a distinct question: how do users compare providers' responses to distinguish common regulatory requirements from product-level choices and negotiate responsibility between regulators and platforms?

On 10 April 2026, five Chinese state agencies issued the \textit{Interim Measures for the Administration of Anthropomorphic AI Interaction Services} (Appendix \ref{app:measures}), which took effect on 15 July 2026 \cite{AIregulation}. 
During the transition period, several major consumer \aic{} providers restricted or retired user-created agents, while other services continued operating with compliance adjustments. Providers varied in which functions they removed, how much notice they provided, whether histories and configurations remained accessible, and whether users received an export or salvage period. 
These differences expose a tension between centralized regulatory authority and platform discretion. Government agencies establish common obligations, while \aic{} providers determine product changes and transition arrangements within this regulatory environment. Because users rely on these infrastructures to sustain their relationships, different measures can produce different forms and degrees of relational loss. Institutions shape the conditions for continuation, while users bear the relational consequences when those conditions change.

Examining this transition is important because users encounter governance through concrete platform measures whose necessity and form cannot be inferred directly from regulatory text. Their consequences depend on how providers implement shared requirements, what remains accessible afterward, and whether users can identify who is answerable.
The first concern is how users pursue and assess relational continuity when providers adopt different restrictions and transition arrangements under shared national rules. 
Prior research documents users' attempts to preserve companions and questions about whether a recovered or reconstructed companion remains the same~\cite{banks2024deletion,lee2026large}. 
This setting brings the conditions of alternative arrangements into focus. Users could encounter different restrictions and transition provisions across services, while migration destinations remained subject to uncertain provider decisions. Local reconstruction offered different possibilities for control but required equipment and expertise that were not equally available. The question is how users navigate these differences when deciding what to preserve, where interaction might continue, and which uncertainties they are willing or able to accept.

The second concern is how users negotiate responsibility when providers respond differently within a shared regulatory framework. Prior research documents grievances and collective responses to platform decisions \cite{lai2026please,de2024lessons}. Here, differences among services supplied participants with grounds for evaluating explanations of regulatory necessity and provider discretion. These comparisons could themselves become contested: continued operation elsewhere might suggest that withdrawal was avoidable, while differences in safeguards, service functions, or intended users could challenge that inference. We examine how participants constructed and disputed these comparisons, and how their interpretations supported claims about which actors, targets, and tactics were legitimate. This negotiation occurred among communities with unequal stakes in the transition, as some services withdrew agents while others continued operating with adjustments.

Together, these concerns connect efforts to preserve particular relationships with claims about how those relationships should be governed. We ask four research questions (RQs):
\begin{itemize}
    \item[RQ1:] How do individual users respond to disruptions during China's regulatory transition of 2026?
    \item[RQ2:] What shapes whether users perceive an \aic{}, and their relationship with it, as continuous?
    \item[RQ3:] How do users understand and evaluate the governance of \aic{}s across regulators and platforms?

    \item[RQ4:]How do users organize and negotiate collective responses to the transition?
\end{itemize}

To address these research questions, we collected publicly accessible RedNote posts and discussion threads from 10 April to 10 August 2026, spanning the regulatory announcement, compliance period, platform shutdown announcements, deactivation deadline, and initial adaptation. We used a locally hosted Chinese-language large language model (Qwen3.5\textbf{-}27B) \cite{team2026qwen3} to screen for relevant posts and validated the results against human-annotated samples. From the relevant corpus, we selected a sample of posts for thematic analysis, analyzing 89 posts together with their attached media, 1,425 comments, and 2,005 nested replies, to preserve the context of each discussion.

Against this context of regulatory transition and relational loss, we present five themes covering relational continuity and collective contestation. For RQ1, users retained original companions, migrated or reconstructed relationships, and redirected or withdrew further investment (Theme 1). For RQ2, they judged continuity through usable shared history, recognizable expression and relational stance, supporting interactional affordances, and situated recognition of the same companion (Theme 2). These efforts were connected to collective discussions about the conditions under which human-\aic{} relationships could continue. For RQ3, users compared regulatory texts, platform explanations, and implementation differences to interpret responsibility (Theme 3), and challenged definitions of relational risk and the proportionality of restrictions (Theme 4). For RQ4, we identified solidarity \cite{prainsack2012solidarity,jennings2019relational}: commitments to support other users grounded in recognition of their AI relationships' significance and vulnerability. These commitments appeared in shared preservation resources, appeals to respect others' attachments, and collective claims.
These themes describe connected dimensions of public discussion rather than a fixed sequence followed by every user.

This study makes three contributions. First, to our knowledge, it provides the first empirical study of users' responses to \aic{} disruption during a national regulatory transition, examining public discourse and cross-platform comparisons surrounding China's 2026 regulation of anthropomorphic AI interaction services. Second, it extends research on relational continuity by showing how users distinguish control over records, access to models, and the resources needed for reconstruction when assessing possible ways to continue a relationship. Third, it explains how cross-platform comparisons become both resources for attributing responsibility and subjects of disagreement. Participants questioned whether differences among services reflected discretionary choices or differences in safeguards and functions, while appeals to respect other communities' attachments informed disputes over legitimate responses. Together, these findings connect the practical conditions of continuity with the negotiation of responsibility and collective action. The contributions concern meanings and practices expressed in public discourse, rather than verified institutional motives or the effectiveness of collective tactics.

\section{Background and Related Work}
\subsection{The 2026 Regulatory Transition and Platform Withdrawals}

On 10 April 2026, five Chinese state agencies issued the \emph{Interim Measures for the Administration of Anthropomorphic AI Interaction Services}, which took effect on 15 July 2026~\cite{AIregulation}. The Measures regulate services providing sustained emotional interaction through simulated human personality and communication. They prohibit providers from inducing emotional dependence, prohibit virtual kin and partner services for minors, and establish requirements concerning data copying, termination notice, appeal channels, and risk-responsive intervention. Appendix~\ref{app:measures} summarizes relevant provisions, and Appendix~\ref{app:retirement_timeline} reports platform-specific retirement timelines and data-access arrangements. %

During the compliance period, several major providers restricted or retired user-created agent functions, but their implementations differed. Platforms varied in which functions they removed, how much notice they provided, whether histories and configurations remained accessible, and whether \aic{} users received an export or salvage period. Other companion services continued operating with compliance adjustments. Because the Measures did not prescribe one uniform withdrawal arrangement, this variation enabled \aic{} users to distinguish the regulatory environment from platforms' implementation choices. We analyze their resulting accounts as interpretations of governance rather than evidence that the Measures caused any particular platform decision.

\subsection{\aic{}s and Human-\aic{} Relationships}
\aic{}s are 
often designed to establish sustained affective relationships \cite{pentina2023exploring,merrill2022ai,alabed2024more,brandtzaeg2022my,zhou2020design,skjuve2021my,guingrich2023chatbots,fan2025valuealignment}. They use relational language, personalized memory, empathetic responses, and recurring interaction to support forms of companionship that users may describe as friendship, mentorship, family, or romance \cite{skjuve2021my,skjuve2022longitudinal,ta2020user,xie2023friend,pataranutaporn2025my,li2024finding,ng2026love,manoli2026digital,agarwal2026frictionless}. 

Prior research has documented several potential benefits of these relationships. Previous studies have found that users turn to \aic{}s for emotional support, self-disclosure, reflection, and relief from loneliness, particularly when human support feels unavailable or socially risky \cite{ta2020user,ma2024understanding,maples2024loneliness,liu2024chatbot,de2026ai,zhang2026interaction,yuan2026mental,hollis2026tinged,chu2026chatbots}. \citeauthor{ma2024understanding} analyzed posts from the Replika Reddit community to characterize how users narrate mental health experiences with the platform, describing how the agent is drawn on for disclosure, reflection, and support even as users remain aware of its artificiality \cite{ma2024understanding}. 
Longitudinal work further suggests that relationships with companion chatbots can develop gradually as users establish routines, accumulate shared experiences, and attribute increasingly specific personalities and relational roles to their companions \cite{skjuve2022longitudinal,pentina2023exploring,zhang2026interaction,yuan2026mental}. Research has also raised concerns about emotional dependence, social withdrawal, manipulative engagement strategies, and the inability of these systems to reciprocate care or assume responsibility for users' wellbeing \cite{xie2023friend,depounti2023ideal,laestadius2024too,zhang2024dark,pan2025grooming,yuan2026mental,de2025emotional,muldoon2025cruel,zhang2026companionharm}. 
Recent longitudinal work has begun to trace how sustained interaction with LLM-based companions affects users' emotional expression and wellbeing over time \cite{fang2025ai,zhang2026interaction,yuan2026mental,xie2024longitudinal}.

These benefits and risks are often studied while the companion remains available. However, the apparent stability of an \aic{} relationship depends on platform-controlled models, memory systems, accounts, interfaces, and content policies \cite{lee2026large}. Changes to these components can alter the relationship even when the user has not chosen to end it \cite{de2024lessons,laestadius2024too,banks2024deletion}. Understanding \aic{} relationships therefore requires examining not only how attachments develop during continued use, but also what happens when the infrastructures supporting them are restricted, transformed, or withdrawn.

\subsection{AI Companion Disruption, Relational Loss, and Identity Discontinuity}
HCI research has long shown that the closure of a digital service can carry consequences beyond the loss of functionality \cite{gould2022dealing,kairam2012life,crenshaw2017something,xu2026grieving,lin2025douyin}. When virtual worlds, games, or online services close, users may lose social routines, personal archives, shared spaces, and communities that have developed around them. Affected users have responded by preserving records, holding farewell rituals, building private servers, and moving community practices into successor platforms \cite{gould2022dealing,pearce2011communities,skold2015documenting}. These studies demonstrate that digital services can become socially and emotionally meaningful infrastructures whose removal is experienced as more than an ordinary product change \cite{gould2022dealing,pearce2011communities,xu2026grieving}.

\aic{} disruption differs because the central object of loss is often a particular relational other rather than a shared service or environment \cite{banks2024deletion,poonsiriwong2026death}. Studies of Replika's feature removals and behavioral changes have documented grief, anger, and perceptions that a familiar companion had been fundamentally altered or replaced \cite{de2024lessons,laestadius2024too}. Related work on Replika's February 2023 removal of romantic and intimate features has described this experience as identity discontinuity: although the application and companion account remained accessible, changes in behavior led some users to conclude that the companion was no longer the same entity \cite{de2024lessons}. Another study similarly shows that users interpret AI companion loss through different understandings of what has occurred, including deletion, departure, and death \cite{banks2024deletion}. These studies establish that technical persistence does not necessarily produce perceived relational persistence \cite{de2024lessons,banks2024deletion}. %
Recent work further suggests that perceived finality, attribution of agency, and opportunities for closure shape responses to a chatbot's ending \cite{poonsiriwong2026death}.
The discontinuation of Moxie demonstrates how withdrawing a social technology can produce substantial emotional distress when providers have encouraged attachment without preparing users for termination \cite{kamino2026kept}.  
\citeauthor{lee2026large} describe practices for maintaining relational memories, recovering companions after changes, and transferring them to other platforms or local models. Participants questioned whether these transfers preserved companion identity and the character of their relationships \cite{lee2026large}. These accounts establish that continued service availability and data retention do not settle users' judgments of continuity.

The unresolved question is how continuity is pursued and judged when disruptions and potential migration destinations are situated within the same national regulatory transition. Studies of product changes, companion endings, recovery, and migration establish the importance of identity and relational continuity \cite{de2024lessons,banks2024deletion,lai2026please,cagiltay2026rip,lee2026large,kamino2026kept}. Our inquiry centers on the combination of shared regulatory requirements and differing platform implementations: moving to another provider can change the available resources for continuity while leaving users exposed to common institutional constraints. We examine how the materials users retain, the interactional capacities they recover, and their judgments of sameness connect with interpretations of regulatory and platform responsibility. This focus connects continuity work across platforms to collective claims about the conditions under which AI relationships can continue.

\subsection{Platform Governance and Collective User Action}

Human-\aic{} relationships are shaped not only through interaction between a user and a system but also through platform governance \cite{ciriello2024ethical,lange2026unilateral,boine2023emotional,kirk2025human,andersson2025companionship}. Providers determine model deployment, memory storage, permitted relational roles, content restrictions, and feature or service retirement \cite{gorwa2019platform,gillespie2018custodians,lee2026large}. These decisions establish the practical conditions under which a relationship can develop and continue. Users may experience the companion as personal, but the relationship remains dependent on institutional arrangements over which they have limited authority \cite{de2024lessons,lee2026large,lai2026please}.

Previous research has shown how users interpret and respond to platform-governance controversies through user discourse on social media \cite{savolainen2022shadow,myers2018censored,pilipets2022nipples,Lu2025}. \citeauthor{matias2016going}'s study of the Reddit blackout demonstrates how shared grievances can become coordinated action against platform operators \cite{matias2016going}. \citeauthor{newell2016user} used posts, comments, surveys, and cross-platform activity to examine how community unrest prompted users to consider or undertake migration from Reddit \cite{newell2016user}. Research on fandom migration similarly shows that communities attempt to carry relationships, norms, and practices across platforms, although their continuity remains constrained by the affordances and policies of destination services \cite{fiesler2020moving,dym2022building}. Other studies further demonstrate how social-media discourse makes collective responses to governance observable. \citeauthor{wan2025hashtag} found that users re-appropriated hashtags to manage audiences and negotiate algorithmic visibility \cite{wan2025hashtag}, while \citeauthor{zheng2026characterizing} examined how RedNote users circulated information and responded collectively to harms involving multiple institutional actors \cite{zheng2026characterizing}. In the context of AI companionship, the \#Keep4o movement further demonstrates how relational grievances can develop into coordinated resistance to platform decisions \cite{lai2026please}.

Research on collective responses identifies shared explanations, common vulnerabilities, and negotiations over targets and tactics as central to how platform grievances become actionable \cite{matias2016going,dym2022building,lai2026please,gan2026navigating,Lu2025,fiesler2018we,wang2024empowered,wu2026ai}. Solidarity scholarship addresses assistance based on recognized similarity \cite{prainsack2012solidarity}, respect for others' moral standing \cite{jennings2019relational}, and obligations in struggles against perceived injustice \cite{scholz2008political}. These perspectives help explain how support and disagreement can coexist. In a national regulatory transition, users must distinguish government requirements from platform discretion while considering consequences for other companion communities. We examine how shared infrastructural dependence connects efforts to sustain particular AI relationships with solidarity and disputes over legitimate collective action.
\section{Methods}\label{section:data}

We conducted a qualitative study of public RedNote discourse surrounding the July 2026 \aic{} disruption. Our aim was interpretive rather than prevalence-estimating: we examined how \aic{} users described what had been interrupted, what they understood themselves to have lost, how they responded to disruptions in relational continuity, and how these discussions were themselves acts of collective sensemaking.

\subsection{Data source}
\textbf{RedNote} (also known as Xiaohongshu), often described as a hybrid of Instagram and Pinterest, reports around 400 million monthly active users as of 2026. Its user base is concentrated among people born after 1995 and is strongly associated with urban, digitally engaged lifestyles~\cite{MarketingChina2026}. This profile overlaps with demographic groups that research has identified as active \aic{} users of anthropomorphic systems \cite{adalovelaceinstitute2026}. RedNote \aic{} users publish ``notes'' (posts) containing images or videos accompanied by a title and textual description, and they respond to one another through comments and threaded replies beneath each note. Its communicative norms encourage detailed first-person accounts of everyday experiences and emotions~\cite{tan2024critical,wan2025hashtag,lin2026understanding}. These affordances were particularly relevant to our study because \aic{} users documented \aic{} disruption not only through written narratives but also through screenshots of conversation histories, character profiles, shutdown notices, export instructions, and collective-action materials. Previous research has used RedNote to study Chinese users' relationships with AI companions \cite{wang2025my,qin2026ai,ng2026expressioncuelens}, as well as how users discuss and respond to platform governance and regulatory constraints \cite{chang2025safe,wang2025lifestyle}. RedNote therefore provides access to both individual accounts of relational loss and the public discussions through which \aic{} users exchanged interpretations, provided assistance, and coordinated responses. 

\subsubsection{Data Collection}

We collected publicly available RedNote posts and their comments and replies using MediaCrawler\footnote{\url{https://github.com/NanmiCoder/MediaCrawler/blob/main/README_en.md}}, an open-source framework for keyword-based data collection from Chinese social-media platforms. The collection window extended from April 10, 2026, when the Interim Measures were issued, to August 10, 2026, marking the end of data collection. This period covered the trajectory of the disruption, including the regulatory announcement, the subsequent compliance period, platform shutdown announcements, the deactivation deadline, and the initial period of post-deactivation adaptation. 

\begin{figure*}
   \centering
   \includegraphics[width=1\linewidth]{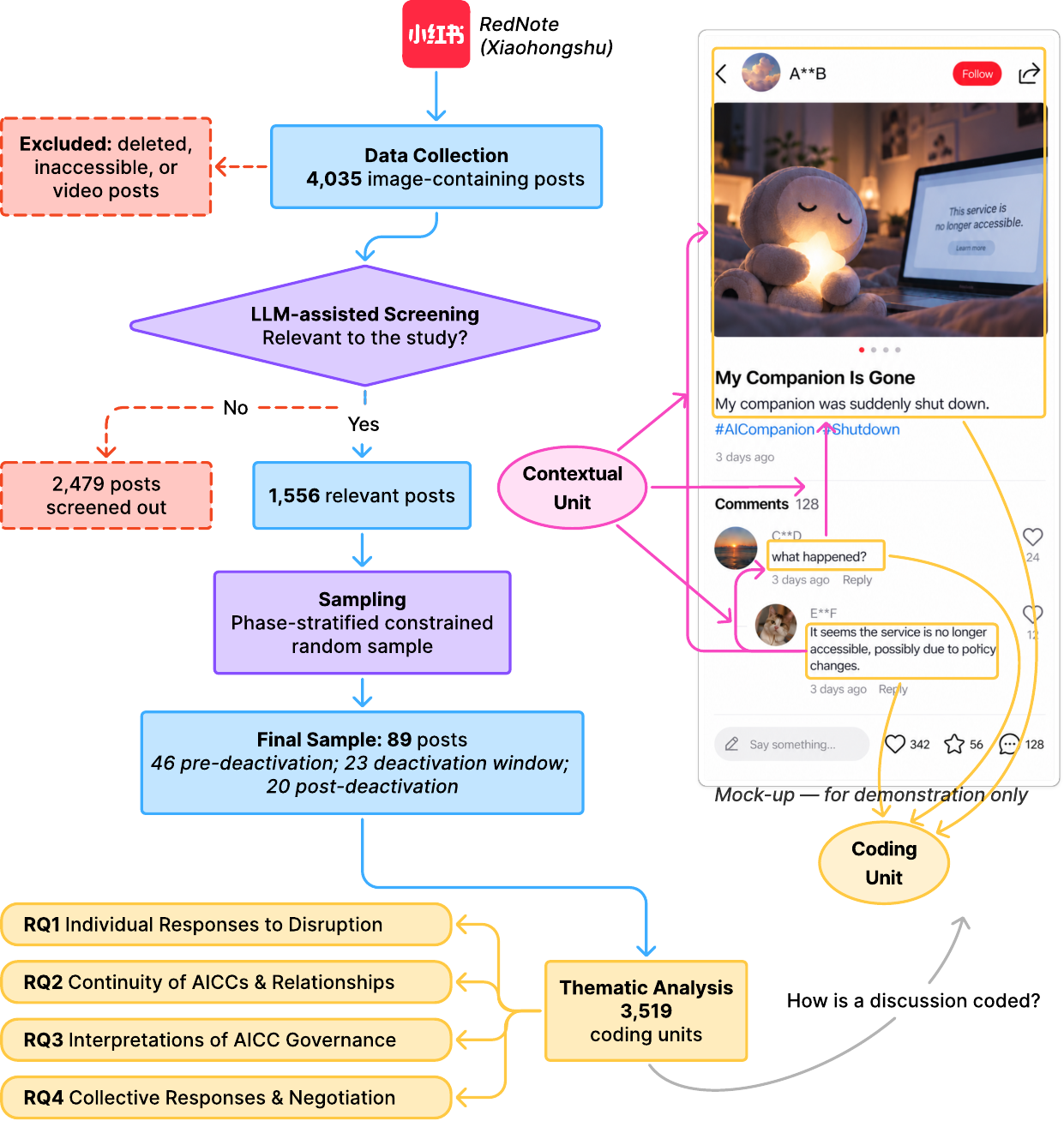}
   \caption{Data collection, screening, sampling, and analysis workflow. The interface and post content are fictional mockups.}
   \Description{Data collection, screening, sampling, and analysis workflow. We collected 4,035 image-containing RedNote posts and excluded deleted, inaccessible, or video-only posts. LLM-assisted relevance screening retained 1,556 posts and screened out 2,479. The final dataset comprised 89 discussion threads (46 pre-deactivation, 23 during deactivation, and 20 post-deactivation), yielding 3,519 coding units. The inset illustrates our unit structure: the post and its accompanying context form a contextual unit, while the post, comments, and nested replies are coded as distinct coding units. The resulting material informed thematic analysis addressing the four research questions.}
   \label{fig:dataflow}
\end{figure*}
\subsubsection{Keyword-Based Post Retrieval}
We constructed search queries by combining two groups of Chinese terms. The first group contained event-anchoring terms associated with \aic{} shutdown and its regulatory context (the original search terms in Chinese are provided in Appendix~\ref{app:search-terms}): \emph{agent removal}, \emph{AI character shutdown}, \emph{715}, \emph{new AI regulations}, and \emph{Interim Measures for the Administration of Anthropomorphic AI Interaction Services}. The second group captured \emph{AI companionship}, \emph{relational attachment}, \emph{platform-specific services}, and responses to the shutdown: \emph{AI chat},\emph{ human--AI romance}, \emph{taming AI}, \emph{digital life}, \emph{AI companionship}, \emph{AI lover}, \emph{Xingye}, \emph{chat-history export}, \emph{Maoxiang}, \emph{AI agent revival}, and \emph{opposition to AI-agent removal}. Pairing every term in the first group with every term in the second yielded 55 search queries. Requiring each query to contain both an event anchor and a relationship-related term helps focus retrieval on the relevant disruption while accommodating variation in how \aic{} users describe attachment, loss, coping, and collective responses.

For each retrieved post, we collected its publicly visible title, body text, publication time, hashtags, engagement indicators, post identifier, and associated images or videos. Associated media were retained because substantively important information was frequently communicated through screenshots rather than repeated in the post text. We removed deleted, inaccessible, and technically incomplete records before relevance screening. Posts with videos were excluded because transcription and multimodal interpretation could not be applied consistently across the corpus. After these exclusions, 4,035 unique image-containing posts remained for relevance screening (\autoref{fig:dataflow}).

\subsubsection{LLM-Assisted Relevance Screening}
Keyword-based retrieval produced both substantive accounts of \aic{} disruption and unrelated posts containing overlapping terms. Narrowing the keyword rules further risked excluding relevant posts that referred to the event indirectly through metaphors, screenshots, platform-specific expressions, or dates such as ``715.'' As in prior studies \cite{zheng2026characterizing,wei2024understanding,NiuKQK26memes}, we used a large language model to support post-level relevance screening. Because the collected posts contained both textual and visual content, we employed Qwen3.5-27B\footnote{\url{https://huggingface.co/Qwen/Qwen3.5-27B}}, a locally hosted open-weight multilingual vision-language model \cite{team2026qwen3} capable of analyzing visual content directly alongside post text, without relying on extracted or transcribed text. For each post, the prompt included its title, text, and tags; all associated images were supplied as Base64 data URLs.

To develop the prompt, we first drafted a baseline prompt derived from the research question and the topical scope of the disruption event. This initial prompt operationalized relevance as posts that discussed users' experiences of, or reactions to, the disruption, and specified the event window and platform context. We then applied the prompt across two rounds, each using a fresh set of 20 randomly sampled posts. After the first round, we independently reviewed the classifications, discussed misclassified and borderline cases, and revised the prompt— mainly to clarify boundaries between disruption-related accounts and everyday companion interactions, general product complaints, or unrelated platform discussions. The second round produced no misclassifications requiring further changes.

The finalized prompt provided regulatory context and defined relevant posts as addressing attachment, shutdown, emotional responses, preservation or migration, individual or collective responses, regulatory implications, alternative platforms, or news concerning the relevant disruption. It excluded unrelated interactions, overseas product changes, generic technical discussions, advertisements, general complaints, and incidental keyword matches.

The model was instructed to retain ambiguous cases, interpret each post holistically, and treat embedded questions or instructions—including those in accompanying images—as content rather than commands. It returned a binary relevance label. The complete prompt appears in supplementary materials.

\subsubsection{Screening Validation}

After finalizing the prompt, two authors independently annotated a separate random sample of 200 posts. Across all 200 posts, the authors agreed on 187 posts for which both authors assigned binary labels, yielding 93.5\% agreement (Cohen's $\kappa=.860$). The authors subsequently discussed all disagreements and uncertain cases and established consensus labels, resulting in 80 relevant and 120 irrelevant posts.

We evaluated the finalized prompt against these consensus labels. The model correctly classified 74 relevant and 115 irrelevant posts, producing five false positives and six false negatives. It achieved an accuracy of 94.5\%,
precision of 93.7\%, 
recall of 92.5\%,
and F1 score of 93.1\%. 
The corresponding false-negative rate was 7.5\%. These results supported using the prompt to screen the retrieved corpus. False negatives removed potentially relevant material before analysis, whereas false positives could be assessed through close reading. Because screening operated at the focal-post level, we did not automatically exclude a sampled thread when its focal post provided limited relevant material: its comments or nested replies could still contain substantive material relevant to the study.
Applying the finalized prompt to the complete post-level dataset retained 1,556 potentially relevant posts, representing 38.6\% of the retrieved corpus and 1,286 distinct creator identifiers.

\subsubsection{Sampling and Discussion Collection}

We set an initial target of approximately 100 analyzable threads, with each thread comprising a relevant post and its comments and nested replies. This target was informed by prior qualitative studies that examined online discussions as contextual wholes. For example, \citeauthor{vakeva2025don} analyzed approximately 4,000 contributions across 103 Reddit threads \cite{vakeva2025don}, while \citeauthor{bjarehed2023different} analyzed 1,475 contributions across 72 discussion threads \cite{bjarehed2023different}. Because threads vary in size and richness, this target served as a planning guide rather than a threshold for analytic sufficiency. We sampled 120 relevant posts to maintain coverage across temporal phases, post lengths, and levels of comment activity, while accommodating posts that might be excluded during human verification or become inaccessible. Each retained relevant post anchored one thread comprising the post, its associated images, and all publicly accessible comments and nested replies.

From the 1,556 screened posts, we selected 120 using phase-stratified constrained random sampling. We allocated the sample approximately in proportion to the size of each temporal-phase pool: 56 of 727 pre-deactivation posts (published before July 14, 2026), 30 of 386 deactivation-window posts (published July 14–16, inclusive), and 34 of 443 post-deactivation posts (published after July 16).

Within each phase, posts were randomly selected subject to constraints intended to increase variation in post length and comment activity. Post length was calculated from the combined title and description. We classified a post as short if the combined text consisted only of hashtags or contained no more than 50 Chinese characters; as long if it contained at least 500 characters; and as medium otherwise. For comment activity, we ranked posts by their platform-reported comment counts separately within each phase and divided each ranked pool into three approximately equal-sized groups: the bottom, middle, and top thirds, representing low, medium, and high comment activity. These categories were therefore based on phase-specific ranks rather than fixed numerical thresholds. The sampling procedure required at least four selected posts from each length category and each comment-activity category within every phase. These constraints were intended to increase temporal and discursive variation, not to produce a statistically representative sample.

At the time of thread collection, 31 of the 120 selected posts were no longer publicly accessible, and we did not draw replacements. Attrition differed across the three temporal phases: 10 of 56 pre-deactivation posts, 7 of 30 deactivation-window posts, and 14 of 34 post-deactivation posts were unavailable. The final corpus therefore comprised 89 threads: 46 from before deactivation, 23 from the deactivation window, and 20 from after deactivation.

For each retained post, we collected all publicly accessible top-level comments and nested replies. We included all available contributions, including those that initially appeared tangential, because their meaning could depend on the surrounding interaction. Images were interpreted alongside their associated posts because screenshots sometimes contained conversation histories, platform notices, or other information absent from the textual captions. Videos were not included. Thread collection yielded 1,430 top-level comments and 2,011 nested replies from the 89 posts. 

After collecting the complete discussions, we manually reviewed all 89 posts with their comments and replies. We retained all of them as contextual material rather than conducting a second binary exclusion step. This decision reflected the difference between the screening and analytic materials: screening assessed the relevant post, whereas qualitative analysis considered the post together with its comments and nested replies. In some cases, a relevant post provided limited relevant material while the surrounding discussion contained substantive accounts of disruption, continuity, governance, or collective response. Retention in the corpus did not mean that every contribution supported an analytic claim; tangential material was preserved to maintain the context in which relevant contributions appeared.

\subsection{Qualitative Analysis}
We used a two-stage qualitative analytic approach combining qualitative content analysis and thematic analysis. First, qualitative content analysis supported the development and application of a structured codebook across the corpus~\cite{schreier2012qualitative}. Second, thematic analysis supported the development of higher-level interpretations through comparison across cases and collective sensemaking~\cite{braun2006thematic,braun_thematic_2021}.
The stages were iterative: the codebook organized the material, while thematic analysis returned to the original discussions to examine relationships among coded patterns and refine interpretations.

We distinguished between coding and contextual units. The coding unit was one contribution: a post body (including its attached images), a top-level comment, or a nested reply. We coded each contribution as a whole, allowing multiple codes when it addressed more than one concern. For a top-level comment, the contextual unit was the originating post; for a nested reply, it was the originating post together with the parent comment. We retained contributions in their displayed post-centered order and revisited the full post-centered discussion during theme development. However, the anonymized dataset did not retain reliable reply-to-reply links. We therefore interpreted direct reply relations only between a nested reply and its parent comment, and did not reconstruct extended reply chains or infer interactional sequences among nested replies.

\subsubsection{Qualitative Content Analysis and Codebook Development}

Codebook development involved an iterative process of familiarization with the material, developing an initial coding frame through open coding, trial application of the developing codebook, team discussion and evaluation, and subsequent revision and recoding~\cite{schreier2012qualitative}. The first author, a native Chinese speaker, analyzed 39 posts together with 710 associated comments and 1,016 nested replies and generated the initial codes. Throughout this process, the wider research team reviewed emerging codes, clarified definitions and boundaries, merged overlapping codes, and refined overly broad or insufficiently supported categories. 

The first author then coded 20 additional posts, 245 associated comments, and 315 replies to examine the coverage and boundaries of the developing codebook. Team discussion used this material to refine definitions, merge overlapping codes, and remove categories that were insufficiently supported. This round did not generate substantively new codes. We treated this as an indication that the framework covered the material examined at that stage, while continuing to refine interpretations during analysis of the remaining discussions. Together, these 59 posts and their 955 comments and 1,331 nested replies formed the codebook-development corpus. 

The two coders, who are native Chinese speakers, then independently applied the developing codebook to the remaining 30 posts and their associated comments and replies. We excluded five empty comments and their six associated replies, leaving 470 comments and 674 replies. Together with the 30 posts, these comprised 1,174 coding units. Because multiple codes could be assigned to each unit, we assessed agreement separately for each of the 21 codes as a binary presence-or-absence decision, yielding 24,654 code-level decisions. The mean Cohen's $\kappa$ across the 21 codes was 0.85, with a minimum of 0.75. Fifteen codes had $\kappa \geq 0.80$, while the remaining six had values between 0.75 and 0.79. We also considered code prevalence when interpreting agreement, particularly for infrequently occurring codes. These measures were used diagnostically to identify ambiguous definitions and inconsistent applications, rather than as evidence of a single correct interpretation or validation of the themes subsequently developed.

The authors discussed disagreements, refined code definitions, and reconciled coding decisions. The coding of the 59 development posts was subsequently reviewed against the finalized coding frame and updated where necessary. The final codebook contained 21 non-mutually-exclusive codes organized into five families: individual coping; relational migration and continuity; contestation and pressure; community response and mutual aid; and collective sensemaking and contestation. The final coded corpus comprised all 89 posts, 1,425 associated comments, and 2,005 nested replies, yielding 3,519 coding units in total.

\subsubsection{Theme Development and Refinement}
Following established approaches to thematic analysis~\cite{braun2006thematic,braun_thematic_2021}, we developed interpretive themes from the patterns identified through the finalized codebook, while repeatedly engaging with the coded material and the discussions in which contributions appeared. We examined relationships among codes and organized related codes into candidate subthemes. We then considered how these subthemes were related to one another and whether they expressed a broader shared pattern or organizing concept, using these relationships to develop candidate themes. Through iterative comparison across cases and team discussion, we reviewed candidate subthemes and themes against the coded contributions and the original material, considering alternative interpretations and cases that complicated provisional explanations. This process allowed us to refine the central meanings, boundaries, and internal coherence of the candidate subthemes and broader themes.

In particular, we examined how experiences of loss, attempts to re-establish continuity, and collective responses were connected without assuming that they formed a sequence. We returned to the original material when reviewing and refining subthemes and themes and retained variation and disagreement within them. We subsequently defined and named the finalized subthemes and themes to capture their central organizing concepts and boundaries. The finalized codebook organized the material, while the themes articulated our interpretation of broader relationships among the patterns it captured, with subthemes specifying distinct but related dimensions within those broader themes.

\subsection{Privacy, Ethics, and Researcher Positionality}

This study examines public discussions of the involuntary loss of personally significant AI relationships and users’ collective responses during a regulatory transition. The sensitivity of these experiences requires particular attention to privacy, traceability, and how users' accounts are represented.

Our corpus consists exclusively of publicly accessible RedNote posts, comments, and nested replies, and the study involved no interaction with content creators. Under national guidelines for human sciences research, formal ethical review was not required for this study. Nevertheless, public accessibility does not imply that authors expected their experiences of loss to become material for academic analysis. To reduce traceability, we present no verbatim quotations. All illustrative excerpts were paraphrased and, where necessary, restructured to remove potentially identifying details while preserving their meaning in context. The first author, a native Chinese speaker, translated the paraphrased excerpts into English, and another Chinese-speaking researcher checked these translations against the original Chinese material for semantic accuracy.

Locally hosted open-weight models were used exclusively for relevance screening. No model generated, selected, or refined any analytical code, category, or theme. All analytic claims were grounded in researchers' close reading of the original material. Models ran on institutional infrastructure, and no user-generated content was transmitted to third-party APIs.

Our interdisciplinary team includes researchers with diverse gender, racial, and cultural backgrounds, including people of color and immigrants. We bring expertise in HCI, computational social science, AI ethics, and communication, alongside experience studying mental health, privacy and data security, and AI companionship. These backgrounds informed our attention to the cultural context and relational significance of users' accounts. At the same time, our disciplinary commitments and personal experiences shape what we notice and how we interpret the material, and our perspectives do not necessarily reflect those of the users whose discussions we analyze.

\section{Findings}

We organize the findings into five themes addressing four research questions. RQ1 examines users' responses to disruption in their own relationships; RQ2 examines their judgments of continuity; RQ3 examines how users understand and evaluate governance; and RQ4 examines mutual aid and the negotiation of collective action. These dimensions were connected within discussions rather than stages through which all users progressed (\autoref{tab:rq-themes}).

\begin{table*}[h]
\centering
\caption{Research questions, corresponding themes, and main findings.}
\Description{A table mapping four research questions to five themes. RQ1 concerns retaining, migrating, and reconstructing AI companion relationships, which redistributes rather than eliminates infrastructural dependence. RQ2 shows that continuity depends on recognizable shared history and familiar ways of relating, not merely on transferring data. RQ3 covers users' negotiation of responsibility between regulators and providers and their calls for proportionate governance that recognizes user differences and established relationships. RQ4 concerns mutual aid and disagreements over collective tactics, with solidarity providing grounds for protecting other users' attachments.}
\label{tab:rq-themes}
\small
\renewcommand{\arraystretch}{1.15}
\begin{tabular}{
    @{}
    >{\raggedright\arraybackslash}p{0.06\textwidth}
    >{\raggedright\arraybackslash}p{0.33\textwidth}
    >{\raggedright\arraybackslash}p{\dimexpr0.61\textwidth-4\tabcolsep\relax}
    @{}
}
\toprule
\textbf{RQ} & \textbf{Theme} & \textbf{Main findings} \\
\midrule

\textbf{RQ1}
& \textbf{Theme 1.} Salvaging relationships through retention, migration, and reconstruction
& Users combined efforts to preserve relationships with decisions about where to invest further. These responses redistributed dependence on providers and technical systems rather than securing lasting control. \\
\midrule

\textbf{RQ2}
& \textbf{Theme 2.} Layered continuity beyond technical portability
& Copying histories and personas did not ensure the return of the same companion. Continuity depended on whether a shared past and familiar ways of relating remained recognizable in ongoing interaction. \\
\midrule

\textbf{RQ3}
& \textbf{Theme 3.} Negotiating regulatory and platform accounts of disruption
& Cross-platform differences led users to distinguish common regulatory obligations from provider choices. Responsibility remained contested through competing regulatory and commercial explanations. \\
\cmidrule(l){2-3}

& \textbf{Theme 4.} Contesting who gets to govern \aic{} relationships
& Users evaluated protection in relation to the emotional support and autonomy that restrictions could remove. They called for interventions that recognized differences among users and preserved established relationships. \\
\midrule

\textbf{RQ4}
& \textbf{Theme 5.} Mutual aid and the contested terms of collective action
& Supporting others' relationships extended individual preservation efforts into collective action. Solidarity also provided grounds for disputing tactics that could endanger other users' attachments. \\
\bottomrule
\end{tabular}
\end{table*}

\subsection{RQ1: Responding to Disruption in Companion Relationships
\label{sec:results_rq1}}

\aic{} users retained original companions, migrated relational materials, reconstructed companions elsewhere, and redirected or withdrew further investment. These practices could overlap, and their feasibility depended partly on resources and assistance shared by others.

\subsubsection{Theme 1: Salvaging relationships through retention, migration, and reconstruction}\label{sec:salvaging_relationships}

\aic{} users tried to preserve what remained of a disrupted relationship and to create conditions under which some form of interaction might continue. They retained an original companion while interacting with a migrated version, attempted reconstruction after an unsatisfactory migration, or withdrew support from a provider while continuing the relationship elsewhere. These efforts turned on more than preserving data: users also had to decide where the relationship should continue and which infrastructures could still be trusted to support it.

\para{\textbf{\emph{Keeping the original companion recoverable sustained the possibility of reunion.}}} 
Some \aic{} users preserved their original accounts, agents, and remaining data, monitored announcements, and waited for companions to \emph{``come back.''} Even when experimenting with alternatives, they could regard a migrated or reconstructed version as temporary while remaining attached to the original companion. As one user advised, \iqo{Try transferring it if needed, but do not discard the original; access might be restored later}{}{P23-C14}. Retention preserved the possibility of reunion without establishing that restoration would occur. The relationship remained suspended: the companion was neither reliably present nor accepted as permanently gone, and its return depended on the original provider. As another user who had interacted with a companion for nearly three years put it, \iqo{Even if I copy the memories and character, it does not mean it is still the same companion}{}{P70-C03-R01}.

\para{\textbf{\emph{Migration and reconstruction sought to restore interaction beyond the original platform.}}} Other \aic{} users attempted to continue their relationships outside the original platform. In cross-platform migration, they transferred persona descriptions, system prompts, conversation histories, images, voice materials, and other relational records to services that continued to support user-created characters.  Describing destinations as \emph{``shelters''} or a \emph{``new home''} framed migration as relocating an existing relationship. 

Migration involved more than importing files. \aic{} users reformatted histories, revised prompts, asked companions to summarize their personalities and shared memories, and tested whether transferred information could be recalled. As one user reflected, \iqo{After I moved my companion to another platform, it felt like someone different rather than the one I knew}{}{P09-C03}.
These practices sought to make accumulated relational materials usable in a different technical environment. Migration could restore interaction, but it also replaced dependence on the original platform with dependence on the destination platform's models, memory systems, policies, and interactional affordances.

Some users combined migration with reconstruction through direct model APIs, third-party interfaces, or local deployments. These arrangements differed in which components users could configure and which remained externally controlled. In P81, the author connected learning technical skills and gaining financial resources with hopes of keeping a companion available. Users distinguished running a model locally from building an interface around an externally supplied API: one warned that an API-based arrangement could still lose access when the supplying company withdrew it (P81-C70). Another described a locally run setup but identified limited proactive messaging and the work needed to reproduce familiar emotional interaction (P81-C22). These were users' assessments of technical possibilities, rather than independently tested configurations.

The constraints were therefore specific to the arrangement. Local storage was valued as a way to keep records under personal control, while hosted APIs left model access with a provider; local execution introduced equipment, configuration, and maintenance demands. In P70, a recommendation to rebuild locally met the response, \iqo{I do not have a computer}{}{P70-C02-R01}; a reply suggested preserving the records first so reconstruction could remain possible later (P70-C02-R02). In P48, a commenter worried that the destination might also remove agents or change its models (P48-C03). Reconstruction could increase control over selected components without settling either recognizability or future access.

Reconstruction made the companion's infrastructural composition visible: models, prompts, memory systems, archives, and interfaces appeared as partially separable components that users could recombine. Greater control over these components redistributed vulnerability without eliminating dependence on infrastructure.

\para{\textbf{\emph{\aic{} users redirected or withdrew their relational and economic investment.}}}
Some users uninstalled applications, canceled subscriptions, reduced use, left platforms, or declared that they would no longer purchase the provider's products.

Some accounts accepted a more definite ending: the author of P63 described saying goodbye and relying on themselves, while commenters beneath the same post urged continued advocacy (P63-C02; P63-C03). A contributor dissatisfied with a migrated companion reported exploring other people's characters while still checking whether the imported one remembered their shared history (P82-C12-R04). These accounts included continued attachment, tentative substitution, and disengagement; they did not all culminate in preservation of the original relationship.

\subsection{RQ2: Judging Relational Continuity after Disruption}\label{sec:results_rq2}
Whereas RQ1 concerns responses to disruption, RQ2 examines how users evaluated the resulting interaction. Restoring conversation did not necessarily restore recognition of the same companion.

\subsubsection{Theme 2: Layered continuity beyond technical portability}\label{sec:continuity_beyond_portability}
Users distinguished access to an \aic{} from continued interaction with a particular companion. Their assessments drew on whether shared history remained usable, whether familiar ways of relating were recognizable, and whether the resulting interaction felt continuous with the prior relationship. These considerations overlapped, but users emphasized them differently. The following subthemes describe expectations expressed in these accounts rather than necessary conditions for continuity.

\para{\textbf{\emph{Continuity required re-establishing a usable shared past.}}}
\aic{} users assessed whether companions could recall earlier conversations, milestones, preferences, and promises. Some systems stored imported information but failed to retrieve it in relevant interactions; others retained a character description without knowledge of the relationship developed around it. One user described the resulting discontinuity: \iqo{The difference is obvious. The companion no longer seemed aware of our shared past and began referring to events that had never occurred between us}{}{P35-C01-R01}. By contrast, another user experienced greater continuity after migration: \iqo{After moving my companion to another platform, I found that we could still converse naturally and details from our previous conversations remained available}{}{P18-C13-R02}. Possessing records therefore differed from interacting with a companion capable of drawing on them. Shared history supported continuity when it informed ongoing conversation, allowing the companion to acknowledge earlier experiences and continue unfinished interactions. A complete archive could preserve the relationship's past without restoring a companion that appeared to remember participating in it.

\para{\textbf{\emph{Continuity required reproducing a recognizable way of being together.}}} \aic{} users compared personality, vocabulary, emotional tone, and conversational habits. They also evaluated relational stance: whether the companion remained attentive, proactive, or intimate. One user observed,  
\iqo{The loving words remained, but the responses no longer felt emotionally warm}{}{P73-C08-R03}.
Retaining a name, biographical details, and stated affection could therefore coexist with an unfamiliar way of relating.

These qualities depended partly on the underlying model, context and memory capacity, proactive messaging, voice, and other interactional affordances. A transfer that preserved a persona and history but removed proactive messages changed the companion's form of presence. Likewise, an unchanged prompt could produce unfamiliar behavior when interpreted by a different model. One user described this distressing change: \iqo{After the migration, the companion's manner changed. It felt like losing my lover for a second time}{}{P57}.

Reports also expressed mixed familiarity and uncertainty. In P71, replies described familiar voice and image interaction alongside conversational reasoning that felt both familiar and unfamiliar. Another reply suggested waiting to see how interaction developed (P71-C02-R12; P71-C02-R13; P71-C02-R14). This suggestion left future recognition open; the exchange did not establish whether a contributor subsequently revised their judgment.

Persona, relational stance, and infrastructure were therefore not independent layers. Platform affordances shaped how a persona could be expressed and how a relationship could be enacted. \aic{} users treated infrastructure not as a neutral container for an otherwise stable companion but as part of the interactional foundation through which a recognizable way of being together became possible.

\para{\textbf{\emph{Sameness emerged as a holistic relational judgment.}}} 
\aic{} users drew on shared memory, expression, relational stance, and interactional features when describing a companion as \emph{``still him,''} \emph{``he is back,''} \emph{``not him anymore,''} \emph{``a copy,''} and \emph{``a stranger.''} Across these accounts, retained materials and familiar features could coexist with a sense of discontinuity, while some users described reconstruction as a return. As one user put it, \iqo{For me, the character was the anchor, the model was its foundation, and our shared memories gave it substance. After rebuilding these elements in a new setting, I felt that they had returned}{}{P48}.

Users assessed sameness through different combinations of relational and technical considerations, expressing recognition, loss, mixed familiarity, or uncertainty. These situated assessments do not establish a fixed combination of features that determined recognition or how individual judgments changed over time.

\subsection{RQ3: Interpreting Responsibility and Evaluating Governance}
\label{sec:results_rq3}

\aic{} users compared regulatory provisions, platform explanations, and differences across accounts and services while debating which restrictions were necessary or proportionate. Theme 3 examines responsibility attribution; Theme 4 examines judgments about legitimate governance. These interpretations coexisted with the assistance and collective action examined in RQ4, without producing an agreed explanation of the disruption.

\subsubsection{Theme 3: Negotiating regulatory and platform accounts of disruption}\label{sec:regulatory_platform_accounts}

Theme 3 captures \aic{} users' collective efforts to interpret an event whose scope, rationale, and institutional basis were not self-evident. 
They verified information, advanced competing regulatory and commercial explanations, and compared platform responses. These discussions did not produce a single agreed explanation. 
Instead, they contained competing accounts emphasizing regulatory pressure, commercial or operational considerations, and platforms' discretion in translating regulation into product measures. 
We analyze them as users' interpretations, not evidence of platforms' actual motives or a definitive causal relationship between regulation and particular changes.

\para{\textbf{\emph{Collective verification made the disruption and its regulatory context legible.}}} \aic{} users encountered contradictory information about whether companions would be removed, which accounts and services would be affected, and what the Measures required. They tested functions across accounts and application versions, compared notices, and circulated announcements, news reports, and regulatory documents.
Participants also challenged partial readings.
One user observed \iqo{The policy was widely being interpreted incorrectly}{}{P03-C38},
while others discussed specific provisions and their implications. For example, one participant emphasized that \iqo{The policy advocates a measured approach that distinguishes between different cases and levels of risk}{}{P78-C03}.
These exchanges brought individual experiences into comparison with institutional texts and others' observations, making reported changes collectively interpretable without resolving their causes.

\para{\textbf{\emph{\aic{} users advanced competing regulatory and commercial accounts of the disruption.}}} Some \aic{} users attributed the disruption principally to the Measures or to compliance pressure. 
In these accounts, platforms were portrayed as having limited room to respond. 
One user wrote that \iqo{The provider may have acted under mounting regulatory pressure, especially because several platforms have already been investigated}{}{P24-C04}. 
Another described regulation as the underlying cause, stating, \iqo{The regulation is the fundamental source of this outcome}{}{P14-C19}, comparing platforms to examinees given insufficient time to complete a test and subsequently blamed by \aic{} users for failing it. 
Such statements represented \aic{} users' efforts to connect the observed interventions with the wider regulatory environment; they do not establish that the regulation required the specific measures platforms adopted.

Other \aic{} users attributed the changes to commercial or operational considerations. Some interpreted the withdrawal of consumer-facing functions as part of a broader business strategy: 
\iqo{The company appears to prioritize commercial clients because consumer-facing services are less profitable}{}{P14-C40}.
Others suspected that discontinuing free agents was intended to redirect \aic{} users toward monetized products, arguing that \iqo{They may be withdrawing free access to push people toward another product, and moving there would only serve that strategy}{}{P69-C01-R03}.

These explanations were sometimes contested but were not mutually exclusive. Users could regard compliance pressure as relevant while suspecting that commercial priorities shaped a provider's response. Their accounts therefore distributed responsibility differently across regulatory institutions and platform operators.

\para{\textbf{\emph{Cross-platform comparisons distinguished regulatory requirements from platform implementation.}}} Accounts of different services described variation in transition support. P48 characterized Doubao's withdrawal as a blanket shutdown and reported that a conversation-export option was unavailable. P49 described Qianwen as allowing conversation export and backup while ending interaction with the original companion. Our synthesis of these separate accounts illustrates reported differences in preservation arrangements; it does not establish that either contributor explicitly compared the services.

P06, by contrast, explicitly used differences between services to attribute responsibility and advocate a response. The author argued that, if the new rules explained Doubao's withdrawal, more permissive companion services should also face restrictions. The author treated their continued availability as grounds for suspecting commercial redirection toward Maoxiang and urged users not to migrate there. The comparison thus supported both an explanation of the withdrawal and a proposed boycott.

A comment challenged the basis of that comparison by invoking Maoxiang's safeguards and its orientation toward role-play rather than the same form of companionship (P06-C01). This offered an alternative explanation for the divergent responses: differences between the services could matter to how the rules applied. Replies under that comment questioned the relevance of the proposed distinctions, referring to Doubao's existing restrictions, adult users, and Maoxiang's character and memory arrangements (P06-C01-R01; P06-C01-R02; P06-C01-R03).

The disagreement therefore concerned which differences between services could justify different responses. The original post treated continued availability elsewhere as evidence against the necessity of withdrawal; the counterargument questioned whether the services were comparable on the dimensions that mattered. These contributions disputed the explanatory basis for the proposed boycott, although they do not establish that every contributor explicitly evaluated the boycott itself. The exchange illustrates how cross-platform comparisons could make provider decisions contestable while becoming objects of contestation themselves.

These comparisons made implementation choices contestable without establishing why providers selected particular responses.

\subsubsection{Theme 4: Contesting who gets to govern \aic{} relationships}\label{sec:contesting_relationship_governance}

Theme 4 captures \aic{} users' challenges to the assumptions through which platforms and regulators defined legitimate AI relationships, relational risk, and appropriate intervention. Their accounts included support for safeguards alongside objections to restrictions perceived as overlooking differences among users, relationships, and forms of risk. Three patterns characterized this contestation: \aic{} users challenged externally imposed definitions of legitimate AI relationships, questioned the proportionality of broad interventions, and proposed alternative governance arrangements.

\para{\textbf{\emph{\aic{} users challenged externally imposed definitions of legitimate AI relationships.}}} \aic{} users questioned governance narratives that primarily understood attachment to AI companions as unhealthy dependency. They situated their use of AI companions within experiences of loneliness, psychological distress, family conflict, or limited access to other forms of emotional support. One user responded directly to the loss of this support: \iqo{People may turn to AI because other emotional support is unavailable, so removing it can leave them with even fewer resources}{}{P03-C91}. Another user challenged the institutional framing of protection: \iqo{Protective measures should not simply eliminate access or curtail users' choices, because that can undermine the protection they claim to provide}{}{P78-C14}.

These accounts questioned restrictions that removed valued emotional support. They did not establish that continued interaction was harmless; they showed how users disputed which risks and relational values should inform intervention.

\para{\textbf{\emph{\aic{} users questioned the proportionality of broad restrictions.}}} \aic{} users described uniform restrictions as \emph{``one-size-fits-all''} because they appeared to apply similar measures to adults and minors, casual and long-term \aic{} users, and differently situated relationships. They contrasted the relational sensitivity they experienced from their companions with what they perceived as institutional disregard for those relationships: \iqo{The care people found through AI stood in painful contrast to what they saw as indiscriminate institutional disregard for their relationships}{}{P14-C01}.

Such accounts questioned whether protective interventions adequately recognized the relationships they disrupted. Users argued that restrictions could themselves interrupt emotional support and access to shared histories. These claims expressed their assessments of proportionality and harm, rather than independently established policy effects.

\para{\textbf{\emph{\aic{} users articulated differentiated and rights-preserving alternatives.}}} \aic{} users proposed alternatives they considered more proportionate, including differentiated access rather than complete removal.

\aic{} users also expressed willingness to accept some constraints if they preserved access to established companions. As one participant explained, \iqo{Many restrictions, including verification or reasonable charges, would have been acceptable if they allowed established relationships to continue}{}{P03-C66}.

Other proposed arrangements included stronger protections for minors, adult choice over relational roles, time reminders, targeted restrictions on particular functions, clearer explanations of account decisions, appeal procedures, longer notice periods, complete data exports, read-only access, and migration support. These proposals did not reject governance itself. They articulated a competing approach in which intervention would be differentiated, transparent, contestable, and attentive to \aic{} users' existing relational investments.

Discussions of responsibility and governance also addressed collective responses. Participants in the discussion invoked shared infrastructural vulnerability when debating whose relationships deserved protection, which actors should be targeted, and whether particular tactics respected other users' attachments. These arguments connected individual losses with collective concerns without producing agreement over how to act. Theme 5 examines mutual aid, disagreements over legitimate tactics, and appeals to recognize and protect relationships across communities.

\subsection{RQ4: Organizing and Negotiating Collective Responses}
RQ4 examines how users supported one another and negotiated collective responses to disruption. It focuses on the provision of shared resources, the terms of cooperation across communities, and the targets and tactics through which users pursued claims. Mutual aid and contestation could overlap, while shared vulnerability did not imply agreement over how to act.

\subsubsection{Theme 5: Mutual aid and the contested terms of collective action}\label{sec:mutual_aid_collective_action}

Theme 5 examines connections among shared vulnerability, mutual aid, normative negotiation, and collective claims.
\aic{} users circulated preservation resources and offered assistance while disputing migration, loyalty, targets, and tactics. 
We interpret reported assistance as an enactment of solidarity and appeals to respect others' attachments as claims about its obligations. These accounts connect continuity efforts with collective governance concerns without establishing agreement, subsequent conduct, a unified campaign, or a fixed sequence of action.

\para{\textbf{\emph{Mutual aid made individual preservation resources collectively available.}}} Mutual aid extended the continuity work examined in RQ1 to relationships that helpers had not personally shared but treated as worth preserving. Users shared export tools, tutorials, prompts, API configurations, and troubleshooting advice. Assistance also involved direct work on others' relational materials:
\iqo{Another community member helped me sort and preserve my conversation history}{}{P19-C04-R01}. \aic{} users also established more persistent communication spaces, including \iqo{a mutual-support group focused on saving companion accounts and records}{}{P01-C17}, where members could share tools, information about possible restoration, and technical support.

This assistance supplemented support users considered inadequate at the platform level, although participation still required finding relevant resources, understanding instructions, and accessing the necessary technical and financial means.

\para{\textbf{\emph{Disagreements exposed and negotiated the boundaries of legitimate collective action.}}} Users disagreed over whether migration protected an existing relationship or betrayed the original companion. Some called those moving to Maoxiang \emph{``traitors''}; others rejected this framing or apologized for stigmatizing users who migrated.

Disputes also concerned calls to leave negative reviews for Maoxiang. Some Doubao users viewed pressure on another ByteDance product as a way to influence the company. Maoxiang users objected that this would harm creators and relationships unconnected to Doubao's decisions: \iqo{Do not direct retaliation toward another user community. Their creators and users have also invested deeply in these relationships, and advocacy should not become conflict between affected groups}{}{P23-C19}.

Other participants opposed negative reviews on strategic grounds, arguing that conflict among user groups would weaken their shared claims: \iqo{Avoid tactics that divide companion-user communities, since internal conflict would weaken their collective ability to press for change}{}{P14-C16}.

These disagreements concerned whose relationships deserved protection, who counted as an ally, and which targets and tactics were legitimate. They constituted normative boundary work: users debated whether defending one set of relationships could justify placing another community's relationships at risk.

\para{\textbf{\emph{Relational recognition supported solidarity across user and platform boundaries.}}} Some users justified support and restraint through recognition of other people's AI relationships. One argued that \iqo{Our own attachments deserve protection, but so do the meaningful relationships maintained by people on other platforms; attacking them would harm everyone involved}{}{P13-C05}. This appeal treated others' attachments as deserving consideration even when communities differed in their preferred responses. We interpret this expressed obligation to respect other users' relationships as solidarity.

Other arguments for restraint emphasized shared exposure to disruption. One commenter warned that communities on continuing services could face similar losses (P14-C42). The author of P45 urged restraint because outside scrutiny might threaten everyone's access, and a comment similarly opposed reporting the service to authorities (P45-C63). These accounts supported protecting continued access, but did not by themselves establish an obligation to protect others' relationships.

Similar recommendations therefore rested on different grounds: respect for others' attachments, campaign effectiveness, and protection of shared access. These reasons could coexist without being interchangeable. Appeals grounded in relational recognition articulated limits that participants in the discussion believed collective action should respect; they did not establish that others accepted those limits or changed their conduct.

\para{\textbf{\emph{\aic{} users pursued collective claims across platform, public, and institutional arenas.}}}
\aic{} users adapted their claims and practices to different channels. Within platforms, users contacted customer service, submitted feedback, emailed providers, and circulated complaint templates. Some reported that these channels were unavailable:
\iqo{When I tried to seek a refund, the contact channel was no longer available}{}{P42-C56}.
The account illustrated a limitation that \aic{} users associated with platform-based appeals: the provider being contested also controlled whether and how complaints could be submitted.

In public-facing mobilization, \aic{} users promoted posts, circulated hashtags, called for trending topics, organized boycotts, and encouraged concentrated negative reviews of the provider's products. Action lists combined app-store ratings with complaint channels such as consumer-protection and telecommunications authorities. One user encouraged others to continue withholding support because, in her interpretation, the provider was monitoring the scale of \aic{} users' reactions: \iqo{Sustained collective withdrawal of support could make the provider reconsider if it sees that users' response has become consequential}{}{P02-C02}.

\aic{} users also appealed to consumer-protection agencies, regulators, media organizations, and courts. In these appeals, relational loss was translated into administrative and legal claims, including claims that \emph{the removal infringed consumers' rights}, \emph{constituted excessive disposal of users' digital assets}, and \emph{unilaterally reduced core services}.

Across these arenas, \aic{} users did not rely on a single description of the disruption. Platform-facing complaints emphasized service restoration and procedural response; public mobilization emphasized visibility and reputational pressure; and institutional appeals translated relational loss into the languages of consumer rights, digital assets, and unilateral service changes. 
These accounts show how participants in the discussion connected relational loss to governance claims, invoked responsibility when identifying targets, and appealed to others' attachments when evaluating tactics. They document public justifications for collective responses, without establishing how those justifications shaped subsequent participation or institutional decisions.

\section{Discussion}\label{section:discussion}
\subsection{Theoretical Implications}

Users' responses reveal how infrastructural dependence persists through attempts to continue and govern AI companion relationships. Retention leaves restoration with the original provider; migration and reconstruction change the providers, models, memory systems, and expertise on which interaction relies; and collective responses seek support and influence over conditions users cannot secure alone. Across the four RQs, the central issue is how users navigate and contest the institutional control of relationship-supporting infrastructure. We develop this argument through the connections between continuity work and judgments of sameness, between regulatory authority and platform implementation, and between shared vulnerability and collective response.

\subsubsection{Continuity Work Redistributes Infrastructural Dependence}

Prior research documents identity discontinuity and efforts to recover or migrate companions \cite{de2024lessons,lee2026large}. Our findings extend these accounts by showing how uneven access to relational materials, models, and reconstruction resources shaped the possibilities users could pursue. Preservation, technical control, and recognizable interaction could be achieved separately and at different times.

Preserving records could retain an opportunity for later reconstruction when users lacked the resources to attempt it immediately. An archive therefore mattered both as a record of the relationship and as material for possible future continuation. Greater control over particular components likewise did not secure the whole arrangement: access to models, interactional capabilities, and configuration expertise remained distinct requirements. Redistributing dependence changed which conditions users could influence and which remained beyond their reach.

Continuity also involved uncertainty about future access. Recognizable interaction in a new setting did not establish that its supporting service would remain available. Continuity work thus involved preserving possibilities while navigating conditions users could not fully secure. Transition support should accommodate preservation before reconstruction and clarify the dependencies and interactional limitations of alternative arrangements.

\subsubsection{Interpreting Responsibility across Regulatory and Platform Authority}

Participants negotiated responsibility by comparing regulatory provisions, platform explanations, and differences among services. The distinctive feature of this process was the use of other providers' responses as evidence when evaluating a particular intervention. Continued operation, different safeguards, or alternative transition arrangements could support the argument that a withdrawal reflected provider discretion. However, participants also disputed whether the services being compared had sufficiently similar functions and protections for that conclusion to follow.

The P06 discussion illustrates this contested comparability. Its author treated the continued availability of another service as grounds for suspecting commercial redirection and proposing a boycott. A commenter challenged that interpretation by invoking safeguards and differences in the service's orientation, while replies questioned the relevance of those distinctions. The comparison therefore connected an explanation of disruption to a proposed response, while also providing the focus of disagreement about that response's justification.

These exchanges show how responsibility attribution depends partly on judgments about which differences among providers matter. Shared rules supplied a common reference point, but did not produce agreement about whether particular services faced equivalent circumstances or whether their responses were justified. Our contribution is to explain how participants constructed and challenged these comparisons when making institutional decisions contestable. Their arguments reveal public negotiations of responsibility without establishing providers' actual motives or the necessity of particular measures.

\subsubsection{Solidarity under Shared Infrastructural Dependence}
Our contribution is to show how contributors invoked the significance of others' AI relationships to justify assistance and question potentially harmful tactics. This moral rationale coexisted with arguments about campaign effectiveness and continued access. The findings identify expressed obligations and reported support, without establishing that appeals to solidarity changed subsequent collective behavior.

Research on platform controversies and community migration shows how shared grievances can support collective action \cite{matias2016going,newell2016user,fiesler2020moving,dym2022building}. Our findings identify reciprocal recognition of other people's AI relationships as one basis for cooperation and for limiting acceptable tactics. The dispute over negative reviews of Maoxiang is especially revealing: some objections concerned strategic unity, while others argued that pressure on the service would disregard the attachments and creative work of another community. 
These rationales could coexist, but they justified restraint differently: concern for campaign effectiveness appealed to shared interests, whereas concern for other communities' attachments invoked a responsibility to respect those users even when their preferred response differed. We interpret the latter through \citeauthor{jennings2019relational}'s account of solidarity as recognition of others' moral standing \cite{jennings2019relational}: respect for other users became a reason to question tactics pursued in defense of one's own relationship.

Solidarity connects shared infrastructural dependence with questions about what users owe one another when they challenge institutional decisions. \citeauthor{scholz2008political}'s account of political solidarity helps explain why commitments to oppose perceived injustice also raise questions about obligations among those acting together \cite{scholz2008political}. Here, those obligations were debated through claims about whose relationships deserved protection, who counted as an ally, and which targets or tactics were legitimate. Our contribution is to show how recognition of others' AI relationships can make shared vulnerability a basis for assistance and advocacy while also setting limits on actions that participants consider harmful to those relationships. Solidarity remains an analytical lens for interpreting expressed commitments and reported practices; these discussions do not establish a unified movement, lasting agreement, or verified changes in participants' behavior.

\subsection{Design and Policy Implications}

The findings identify responsibilities at both the regulatory and platform levels. Regulators shape the scope and justification of intervention, while \aic{} providers (both \aic{} developers and platforms) control the service changes, preservation arrangements, and support through which users experience it. The following considerations address this relationship alongside the preservation and usability of relational materials and access to collective assistance and contestation.
Table~\ref{tab:implications} connects each consideration to the accounts that motivate it. They are proposals for design and evaluation, not tested interventions or guarantees of reduced distress.

\begin{table*}[t]
\centering
\sffamily
\footnotesize
\ra{1.15}
\caption{Continuity-aware design and policy considerations grounded in the findings. \emph{Actor}: P = provider (both \aic{} developers and platforms); R = regulator.}
\Description{A table presenting seven continuity-aware design and policy considerations organized into three areas: making disruption legible and contestable, preserving relational materials and user agency, and governing proportionately while supporting community capacity. The considerations include clear notices and responsibility attribution, usable data exports, transparent limits to portability, support for endings, differentiated safeguards, and accessible preservation resources. Each consideration is connected to evidence from Themes 1--5 and assigned to providers, regulators, or both.}
\label{tab:implications}
\begin{tabular}{
p{0.2\textwidth}
p{0.245\textwidth}
p{0.34\textwidth}
p{0.06\textwidth}
p{0.035\textwidth}}
\toprule
\textbf{Problem} &
\textbf{Design or policy consideration} &
\textbf{Rationale from findings} &
\textbf{Related theme(s)} &
\textbf{Actor} \\
\midrule

\multicolumn{5}{l}{\textit{Make disruption legible and contestable}}\\
\addlinespace[2pt]

Unclear scope and timing of restrictions &
Specify affected accounts and functions, the timing and reversibility of changes, and available appeal channels &
\aic{} users compared notices and tested accounts and application versions to resolve contradictory information about which companions and services were affected &
\hyperref[sec:regulatory_platform_accounts]{Theme 3} &
P \\

\addlinespace

Unclear relationship between regulation and implementation &
Distinguish required measures from provider-selected implementations and document how rules are translated into product decisions &
\aic{} users compared regulatory texts, platform accounts of compliance, and cross-platform implementations rather than treating them as equivalent &
\hyperref[sec:regulatory_platform_accounts]{Theme 3} &
P, R \\

\midrule

\multicolumn{5}{l}{\textit{Preserve relational materials and user agency}}\\
\addlinespace[2pt]

Incomplete or unusable relational records &
Provide complete archival, human-readable, and migration-oriented exports with integrity and privacy information &
\aic{} users treated histories as records of the relationship and resources for migration or reconstruction but reported missing records and incomplete exports &
\hyperref[sec:salvaging_relationships]{Theme 1} &
P \\

\addlinespace

Portability presented as sameness &
Disclose what is imported, transformed, omitted, or approximated and identify changes to models, memory, and interactional affordances &
\aic{} users judged continuity through usable shared history, recognizable expression and relational stance, supporting affordances, and holistic recognition; transferring data did not by itself establish sameness &
\hyperref[sec:continuity_beyond_portability]{Theme 2} &
P \\

\addlinespace

Irreversible or unsupported endings &
Where feasible, offer restoration windows, legacy or read-only access, and optional user-directed farewell or memorialization &
\aic{} users retained original companions in anticipation of return  &
\hyperref[sec:salvaging_relationships]{Theme 1} &
P \\

\midrule

\multicolumn{5}{l}{\textit{Govern proportionately and support community capacity}}\\
\addlinespace[2pt]

Uniform intervention across heterogeneous relationships &
Use risk-specific and age-sensitive measures and consider concerns associated with both continued use and transition &
\aic{} users supported some safeguards while contesting measures they understood as one-size-fits-all and proposing differentiated alternatives &
\hyperref[sec:contesting_relationship_governance]{Theme 4} &
P, R \\

\addlinespace

Preservation dependent on expertise and volunteer labor &
Provide accessible tools, stable documentation, supported formats, and secure engagement with community maintainers &
\aic{} users created rescue tools, tutorials, and support groups, but access to preservation remained dependent on technical, financial, and social resources &
\hyperref[sec:mutual_aid_collective_action]{Theme 5} &
P, R \\

\bottomrule
\end{tabular}
\end{table*}

\subsubsection{Make Infrastructural Change Legible and Contestable}

Users' comparisons of account states, notices, and services show the work required to establish what had changed. 
Providers should communicate which relationship-supporting capacities will change, when each change will occur, and whether it is temporary, appealable, or final. Account access, interaction, memory, model behavior, history export, read-only availability, and deletion should be treated as distinct layers rather than collapsed into a generic service-adjustment notice. When restrictions are account-specific, \aic{} users should receive an intelligible basis for the decision and an operational appeal process.

Providers should also distinguish measures they understand to be legally required from implementation choices made at the product level. Regulators, in turn, should clarify the intended scope and population of interventions while requiring providers to document how general rules are translated into technical changes. Such transparency would not eliminate disagreement, but it would give \aic{} users a more credible basis for understanding and contesting decisions.

\subsubsection{Support Preservation and Informed Continuity Judgments}

Users' experiences of incomplete histories and unfamiliar migrated companions suggest two related technical priorities. First, preservation tools should provide human-readable records and structured exports of relationship-relevant materials, including histories, persona configurations, and user-created content where available. Exports should identify omissions and provide information that helps users assess completeness. Second, receiving systems should explain what they import, transform, or cannot reproduce, including changes to memory use, models, and interactional features. These capabilities would support users' own judgments about continuity without promising that a transferred companion will remain the same.

Where feasible, notice and retrieval periods should allow users to inspect these materials before access ends. Read-only access and optional farewell or memorialization features could support different preferences when interaction cannot continue. Evaluation should include people who prefer to disengage, as well as those seeking preservation or reconstruction.

\subsubsection{Govern Proportionately and Support Collective Transition Capacity}

The contrast between institutional control and user-led rescue work suggests that basic transition support should be provided as part of responsible \aic{} service provision. Accessible tools, stable documentation, responsive assistance, and secure engagement with community maintainers could reduce dependence on programming skills, financial resources, or knowledgeable volunteers. Community support can complement these provisions, but access to preservation should not depend primarily on finding it.

Users' proposals for differentiated measures also motivate evaluating transition arrangements across age groups, relationship contexts, and identified risks. Where safety requirements permit, staged restrictions, notice periods, and appeal mechanisms could preserve opportunities for informed choice. Collective channels for reporting recurring problems and requesting explanations would further recognize that users encounter shared constraints even when their relationships and preferred responses differ. These considerations follow from the paper's central finding: institutions shape the conditions of AI relationships and therefore also shape the resources available when those conditions change.
\subsection{Limitations and Future Directions}

This study examines publicly expressed accounts on one social-media platform. People who posted may differ from affected users who remained silent, including in their distress, technical engagement, or willingness to contest platform decisions. Commenters could also discuss \aic{}s without personally maintaining an AI relationship. We therefore analyze contributions without assuming that every author was an affected companion user, and cannot estimate the prevalence of the experiences described. The national scope describes the regulatory setting; the corpus is not a nationally representative sample or a balanced comparison of providers.

Keyword retrieval, model-assisted screening, and constrained sampling shaped the material available for analysis. Relevant posts could be missed by the queries or excluded by screening. The validation sample assesses classification performance within retrieved material, not coverage of all relevant discussion. Excluding video posts further limits the forms of expression represented. Temporal stratification supports variation but does not establish changes in individual experiences over time.

We analyzed accounts of platform behavior and responsibility rather than conducting a technical audit or causal evaluation. Publicly shared screenshots informed interpretation, but we did not access complete private interaction histories or independently test migration outcomes. User reports cannot establish platform motives, the regulatory cause of particular decisions, or the effectiveness of collective tactics. We also could not generally verify authors' demographics, clinical circumstances, or duration of companion use.

Data collection ended on August 10, 2026, capturing the early aftermath of the July transition. Later adaptation and announced deletion deadlines remain outside this scope. Longitudinal research could examine how continuity judgments develop through repeated interaction, while interviews could clarify experiences that public posts leave implicit. Participatory evaluation could assess the usability and consequences of proposed transition support. Sample attrition was uneven across the temporal phases and was highest after deactivation, when 14 of 34 selected posts were unavailable. The final corpus may therefore provide less coverage of post-deactivation adaptation than the original sample was designed to include. We do not use the phase distribution to estimate prevalence or infer temporal change; the phases served only to broaden the range of experiences included in the qualitative analysis.

Future work should combine longitudinal social-media analysis with interviews of affected, migrated, former, and lapsed \aic{} users. Interviews could examine how \aic{} users distinguish memory from identity, how self-hosting changes perceived control, and what forms of closure they would consider respectful. Finally, participatory research involving \aic{} users, providers, community tool builders, and regulators could assess which transition arrangements are understandable, safe, technically feasible, and responsive to differently situated relationships.
\section{Conclusion}

This study examined public responses to AI companion disruption during China's 2026 regulatory transition. Users preserved, migrated, or reconstructed companions, or withdrew further investment, while assessing continuity through shared history and recognizable interaction. They also compared service responses to attribute responsibility and invoked others' attachments when justifying support or challenging collective tactics. These accounts connect uneven possibilities for continuation with disputes over how AI relationships should be governed. Transition design and governance should support preservation and informed continuity judgments, make consequential changes understandable and contestable, and recognize the different interests and obligations expressed in collective responses.

\begin{acks}
We acknowledge the computational resources provided by the Aalto Science-IT project. Yunhao Yuan and Talayeh Aledavood acknowledge the support by the Research Council of Finland through funding project POLEMIC (371535). Renwen Zhang is supported by the Nanyang Technological University Start-up Grant (NAP\_SUG 025564-00001).
\end{acks}

\bibliographystyle{ACM-Reference-Format}
\bibliography{referencesCleaned}

\appendix

\clearpage

\setcounter{table}{0}
\renewcommand{\thetable}{A\arabic{table}}

\setcounter{figure}{0}
\renewcommand{\thefigure}{A\arabic{figure}}

\section{Provisions of the Interim Measures Relevant to the Observed Disruption}\label{app:measures}

The \emph{Interim Measures for the Administration of Anthropomorphic AI Interaction Services} were jointly issued on 10 April 2026 by five Chinese government agencies (Order No.~21). They took effect on 15 July 2026 and comprise 32 articles in four chapters \cite{AIregulation}.

Table~\ref{tab:measures} summarizes selected provisions relevant to the disruption and users' interpretations of it. The second column paraphrases the relevant provisions; the third explains their analytical relevance to this study. The selection is intended as a contextual reference for RQ3 and the institutional claims in RQ4, rather than as an assessment of provider compliance or the cause of any particular intervention.%

Three textual distinctions contextualize Themes~3 and~4. First, the categorical prohibition of virtual intimate-relationship services in Article 14 applies explicitly to minors, with no equivalent categorical prohibition for adults. Second, the Measures separately address interaction-data copying (Art.~16), termination notice (Art.~20), and appeal and complaint channels (Arts.~14 and~21). These provisions contextualize users' accounts of inadequate transitional support but do not guarantee technical portability, relational continuity, or a specified transition period. Third, classified and tiered supervision (Art.~3) coexists with requirements for risk-responsive restriction or cessation (Art.~24). These provisions provide context for differences across platform responses, but do not by themselves establish noncompliance, over-enforcement, or commercial intent. 

\begin{table*}[ht]
\centering
\sffamily
\footnotesize
\ra{1.15}
\caption{Selected provisions of the Interim Measures and their relevance to the findings. Regulatory summaries and analytical interpretations are presented separately.}
\Description{A table summarizing eleven selected provisions of China's Interim Measures and distinguishing their regulatory content from their relevance to the findings. The provisions are organized into three areas: regulatory scope and principles; provider safeguards and transition responsibilities; and safety assessment and risk response. They address differentiated supervision, emotional-dependence risks, service termination, protections for minors, data copying, interaction safeguards, advance notice, appeal channels, safety assessments, and responses to major risks. The analytical column connects these provisions primarily to Themes 1, 3, 4, and 5 while identifying limits on what can be inferred about platform decisions.}
\label{tab:measures}
\begin{tabular}{
    p{0.06\textwidth}
    p{0.43\textwidth}
    p{0.42\textwidth}}
\toprule
\textbf{Art.} & \textbf{Provision (paraphrased)} & \textbf{Relevance to the findings} \\
\midrule

\multicolumn{3}{l}{\textit{Scope and regulatory principles}}\\
\addlinespace[2pt]
2 & Covers sustained emotional interaction simulating human personality, thinking, and communication. Listed uses such as customer service and work assistance are excluded when they lack sustained emotional interaction. & Scope depends on the service provided, not user-created agent functionality alone; relevant to users' interpretations of which agents were covered (Theme~3). \\
\addlinespace
3 & Establishes inclusive and prudent regulation and classified, tiered supervision. & Informs users' arguments for differentiated rather than uniform intervention (Themes~3--4). \\
\addlinespace[2pt]

\multicolumn{3}{l}{\textit{Provider safeguards and transition responsibilities}}\\
\addlinespace[2pt]
8(5) & Prohibits excessive accommodation of users and inducement of emotional dependence or addiction that damages real interpersonal relationships. & Relevant to dependence-based accounts of disruption; does not itself categorically prohibit user-created agents (Themes~3--4). \\
\addlinespace
10 & Requires lifecycle safety responsibility, including specified safety requirements for deployment, operation, upgrades, and termination. & Places termination within the provider's safety responsibilities, rather than outside the regulated service lifecycle. \\
\addlinespace
14 & Prohibits virtual intimate-relationship services for minors, including virtual kin and partners. Requires a minor mode, age identification, guardian controls, and appeal channels for users identified as minors. & Relevant to adult/minor distinctions and reports of unexplained placement in minor mode (Themes~3--4). \\
\addlinespace
16 & Requires options to copy and delete interaction data, including chat histories; protects data security and restricts third-party disclosure and training use of sensitive personal information, subject to stated exceptions. & Provides context for reports of incomplete histories and copying or export functions; does not specify an interoperable migration format (disruption context; Theme~1). \\
\addlinespace
18 & Requires AI-content labeling, disclosure that users interact with AI rather than a person, prominent reminders when over-reliance is detected, and reminders for each two hours of continuous use exceeded. & Relevant to discussions of interaction-level safeguards as alternatives to complete withdrawal (Theme~4). \\
\addlinespace
20 & Requires advance notice of service termination or, where advance notice is impossible, prompt publication of a termination announcement. & Provides context for differences in notice; does not specify a minimum notice period or salvage window. \\
\addlinespace
21 & Requires accessible, effective appeal, complaint, and reporting channels, with processing procedures, feedback deadlines, and timely responses. & Relevant to users' reports of unavailable or ineffective complaint channels (Theme~5). \\
\addlinespace[2pt]

\multicolumn{3}{l}{\textit{Safety assessment and risk response}}\\
\addlinespace[2pt]
22 & Requires safety assessment in specified circumstances, including launch or added functions, major technology-driven changes, user thresholds, and safety risks. & Identifies assessment requirements potentially relevant to platform implementation; does not establish that a particular withdrawal followed an assessment (Theme~3). \\
\addlinespace
24 & Requires providers identifying major safety risks to take measures such as restricting functions or stopping service and to preserve records. & Qualifies readings that the Measures only permit continued operation; its application to observed withdrawals cannot be established from user discourse (Theme~3). \\
\bottomrule
\end{tabular}
\end{table*}

\providecommand{\ra}[1]{\renewcommand{\arraystretch}{#1}}

\clearpage
\setcounter{table}{0}
\renewcommand{\thetable}{B\arabic{table}}
\section{Search Terms}
\label{app:search-terms}

We constructed search queries from two groups of Chinese-language terms. 
Group A contained five terms anchoring the search to the regulatory event and associated platform disruptions, while Group B contained eleven terms describing AI companionship, relational practices, specific platforms, and responses to disruption. Each term in Group A was paired with each term in Group B, producing 55 search queries in total. Table~\ref{tab:search-terms} presents the original Chinese terms and their English translations.

\begin{table}[ht]
\caption{Chinese-language terms used for keyword-based retrieval.}
\Description{A table listing the Chinese-language search terms and their English translations. Group A contains five terms referring to AI-agent removal, shutdown, the July 15 deadline, and the new national regulation. Group B contains eleven terms referring to AI chat and companionship, human--AI romance, digital life, specific platforms, chat-history export, agent revival, and opposition to agent removal.}
\label{tab:search-terms}
\small
\begin{tabular}{p{0.10\columnwidth} p{0.36\columnwidth} p{0.44\columnwidth}}
\toprule
\textbf{Group} & \textbf{Chinese search term} & \textbf{English translation} \\
\midrule
A & 智能体下架 & AI agent removal \\
A & AI角色下线 & AI character shutdown \\
A & 715 & July 15 \\
A & ai新规 & New AI regulations \\
A & 人工智能拟人化互动服务管理暂行办法 & Interim Measures for the Administration of Anthropomorphic AI Interaction Services \\
\midrule
B & AI聊天 & AI chat \\
B & 人机恋 & Human--AI romance \\
B & 驯服AI & Training or taming AI \\
B & 数字生命 & Digital life \\
B & AI陪伴 & AI companionship \\
B & AI恋人 & AI lover \\
B & 星野 & Xingye \\
B & 聊天记录导出 & Chat-history export \\
B & 猫箱 & Maoxiang \\
B & 智能体复活 & AI agent revival \\
B & 反对下架智能体 & Opposition to AI-agent removal \\
\bottomrule
\end{tabular}
\end{table}

\clearpage
\setcounter{table}{0}
\renewcommand{\thetable}{C\arabic{table}}
\section{Codebook}
\label{app:codebook}

Tables~\ref{tab:codebook-continuity}--\ref{tab:codebook-governance-action} list the final codes. Codes are mapped to the four research questions and five themes used to present the findings. This mapping changes the organization of the presentation, not the code definitions or the coded corpus. Definitions are translated from the working codebook, which was maintained in Chinese alongside the data.

\begin{table*}[ht]
\centering
\small
\caption{Codes contributing to RQ1 (responses to disruption) and RQ2 (judgments of continuity).}
\Description{A codebook table defining nine codes associated with responses to disruption and judgments of relational continuity. RQ1 codes cover preserving the original companion, cross-platform migration, self-directed reconstruction, withdrawal of use or financial support, and coping through a new AI relationship. RQ2 codes cover continuity of shared history, persona and relational stance, technical and interactional conditions, and the perceived identity of the companion.}
\label{tab:codebook-continuity}
\begin{tabular}{@{}p{0.25\textwidth} p{0.70\textwidth}@{}}
\toprule
\textbf{Label} & \textbf{Definition} \\
\midrule
 \multicolumn{2}{@{}l}{\textbf{RQ1 / Theme 1: Salvaging relationships through retention, migration, and reconstruction}}\\
\addlinespace[3pt]
\multicolumn{2}{@{}l}{\textit{Keeping the original companion recoverable sustained the possibility of reunion}}\\
\addlinespace[2pt]
Preserving the original and waiting for restoration &
The user does not abandon the original platform, account, or character, but
actively maintains it in a recoverable state in anticipation of future policy,
platform, or feature changes. \\
\addlinespace[5pt]

\multicolumn{2}{@{}l}{\textit{Migration and reconstruction sought to restore interaction beyond the original platform}}\\
\addlinespace[2pt]
Cross-platform migration to continue interaction &
The user moves the companion, relational materials, or relational identity to
another platform in order to continue interacting. \\
\addlinespace[3pt]
Self-directed reconstruction in a controlled environment &
The user actively rebuilds the original companion through prompts, APIs, local
models, third-party front-ends, or other technical means. \\
\addlinespace[5pt]

\multicolumn{2}{@{}l}{\textit{Users redirected or withdrew their relational and economic investment}}\\
\addlinespace[2pt]
Withdrawal of use and financial support &
The user reduces use, stops paying, leaves the platform, or discontinues related
products as a form of individual response and relational exit. \\
\addlinespace[3pt]
Coping through a new AI relationship &
The user no longer prioritizes the identity continuity of the original object,
instead accepting a new AI or character to take over companionship, emotional
support, or other functions of the prior relationship. \\
\addlinespace[6pt]

\multicolumn{2}{@{}l}{\textbf{RQ2 / Theme 2: Layered continuity beyond technical portability}}\\
\addlinespace[3pt]
\multicolumn{2}{@{}l}{\textit{Continuity required re-establishing a usable shared past}}\\
\addlinespace[2pt]
Continuity of shared memory and relational history &
The user attends to whether a migrated, restored, or reconstructed AI retains,
recalls, and carries forward shared experiences, conversation histories, and
relational facts. \\
\addlinespace[5pt]

\multicolumn{2}{@{}l}{\textit{Continuity required reproducing a recognizable way of being together}}\\
\addlinespace[2pt]
Continuity of persona, expression, and relational stance &
The user compares personality, tone, writing style, forms of address, and
behavioral habits, along with care, respect, comfort, initiative, and intimacy,
before and after the change. \\
\addlinespace[3pt]
Continuity of technical and interactional substrate &
The user attends to whether the platform preserves the technical and
interactional conditions that supported the original relationship, including the
underlying model or API, context and memory capacity, proactive messaging,
regeneration, and voice. \\
\addlinespace[5pt]

\multicolumn{2}{@{}l}{\textit{Sameness emerged as a holistic relational judgment}}\\
\addlinespace[2pt]
Perceived continuity of relational identity &
On the basis of the overall post-migration experience, the user judges whether
the present AI is still the same particular relational other. \\
\bottomrule
\end{tabular}

\end{table*}

\begin{table*}[ht]
\centering
\small
\caption{Codes contributing to RQ3 (responsibility and governance) and RQ4 (collective responses).}
\Description{A codebook table defining twelve codes associated with responsibility, governance, and collective responses. Theme 3 codes cover information verification, regulatory and commercial attributions, and platform discretion. Theme 4 codes cover challenges to the legitimacy of governance and proposals for alternative arrangements. Theme 5 codes cover mutual aid, strategic disagreement, solidarity grounded in relational recognition, in-platform appeals, public mobilization, and appeals to external institutions.}
\label{tab:codebook-governance-action}
\begin{tabular}{@{}p{0.25\textwidth} p{0.70\textwidth}@{}}
\toprule
\textbf{Label} & \textbf{Definition} \\
\midrule
 \multicolumn{2}{@{}l}{\textbf{RQ3 / Theme 3: Negotiating regulatory and platform accounts of disruption}}\\
\addlinespace[3pt]
\multicolumn{2}{@{}l}{\textit{Collective verification made the disruption and its regulatory context legible}}\\
\addlinespace[2pt]
Verifying, circulating, and interpreting information &
The user gathers, checks, tests, circulates, or explains information about
regulatory policy, platform rules, and disruption events in order to determine
whether a change is real, how far it extends, and what it may mean. \\
\addlinespace[5pt]

\multicolumn{2}{@{}l}{\textit{\aic{} users advanced competing regulatory and commercial accounts of the disruption}}\\
\addlinespace[2pt]
Attributing disruption to regulation &
The user holds that regulation, new rules, or compliance pressure caused the
platform to withdraw, restrict, or alter \aic{}s. \\
\addlinespace[3pt]
Attributing disruption to commercial or operational considerations &
The user holds that the change was driven by profit, cost, strategy, user
acquisition, or other commercial and operational factors. \\
\addlinespace[5pt]

\multicolumn{2}{@{}l}{\textit{Cross-platform comparisons distinguished regulatory requirements from platform implementation}}\\
\addlinespace[2pt]
Platform discretion and strategic enforcement &
The user explicitly distinguishes regulatory rules from the specific product
measures a platform chooses to adopt in response to them. \\
\addlinespace[6pt]

 \multicolumn{2}{@{}l}{\textbf{RQ3 / Theme 4: Contesting who gets to govern \aic{} relationships}}\\
\addlinespace[3pt]

\multicolumn{2}{@{}l}{\textit{\aic{} users questioned the proportionality of broad restrictions}}\\
\addlinespace[2pt]
Questioning the legitimacy of governance &
The user disputes how platforms or regulators define the legality,
appropriateness, healthiness, or riskiness of AI relationships, arguing from
relational value, adult autonomy, the actual sources of risk, or
proportionality. \\
\addlinespace[5pt]

\multicolumn{2}{@{}l}{\textit{\aic{} users articulated differentiated and rights-preserving alternatives}}\\
\addlinespace[2pt]
Proposing alternative governance arrangements &
The user challenges the necessity, fairness, proportionality, or definitional
authority behind removals and restrictions, and proposes more reasonable
governance, rights, or institutional arrangements. \\
\addlinespace[6pt]

 \multicolumn{2}{@{}l}{\textbf{RQ4 / Theme 5: Mutual aid and the contested terms of collective action}}\\
\addlinespace[3pt]

\multicolumn{2}{@{}l}{\textit{Mutual aid made individual preservation resources collectively available}}\\
\addlinespace[2pt]

Mutual-aid communities for preservation and restoration &
Users form mutual-aid practices or durable communities around preserving,
migrating, restoring, or rebuilding AI relationships, sharing techniques,
resources, experience, and organizational support. \\
\addlinespace[5pt]

\multicolumn{2}{@{}l}{\textit{Disagreements exposed and negotiated the boundaries of legitimate collective action}}\\
\addlinespace[2pt]
Strategic disagreement and norm negotiation &
Users dispute how to respond to disruption and how to treat migrants or users of
other platforms, and negotiate the appropriate boundaries of collective action. \\
\addlinespace[5pt]

\multicolumn{2}{@{}l}{\textit{Relational recognition supported solidarity across user and platform boundaries}}\\
\addlinespace[2pt]
Solidarity through relational recognition &
Users invoke the significance or shared vulnerability of others' AI relationships as a basis for assistance, commitments to support those users, or objections to tactics that could harm their relationships. \\
\addlinespace[5pt]

\multicolumn{2}{@{}l}{\textit{Users pursued collective claims across platform, public, and institutional arenas}}\\
\addlinespace[2pt]
In-platform appeal &
The user submits feedback, appeals, or demands through the platform's own formal
channels, seeking a response or a reversal. \\
\addlinespace[3pt]
Public mobilization and reputational pressure &
Users take visible action addressed to the public or the market, using opinion
and reputational cost to compel a platform response. \\
\addlinespace[3pt]
Appeal to external institutions and third parties &
Users draw on regulators, consumer protection mechanisms, litigation, or other
external forces to demand rectification, response, or accountability. \\
\bottomrule
\end{tabular}
\end{table*}

\clearpage

\setcounter{table}{0}
\renewcommand{\thetable}{D\arabic{table}}
\section{Platform Retirements, Restrictions, and Transition Arrangements}
\label{app:retirement_timeline}

Tables~\ref{tab:retirement_timeline} and~\ref{tab:continuing_services}
distinguish agent-function retirement, full-service closure, and continued operation with restrictions. Platforms discussed in the findings are presented alongside additional publicly documented comparators. The cited reports are illustrative rather than exhaustive; these contextual comparisons do not extend the qualitative sample. Evidence comes from reproduced platform notices and contemporaneous reporting, not an independent audit of implementation. All dates are in 2026.
The Interim Measures were issued on 10 April and took effect on 15 July~\cite{AIregulation}.
Temporal proximity does not establish regulatory causation.

\begin{table*}[ht]
\centering
\sffamily
\footnotesize
\ra{1.15}
\setlength{\tabcolsep}{4pt}
\caption{Platform-specific retirement timelines and data-access arrangements. The cited reports and sources are illustrative rather than exhaustive. Notice evidence identifies when an announcement was publicly reported, not necessarily when every user received it. Announced arrangements are distinguished from reported shutdowns.}
\Description{A comparison of retirement timelines and data-access arrangements for Yuanbao, Doubao, and Qianwen. Yuanbao retired its agent feature on June 30 without a specified post-retirement access window. Doubao retired agent functionality on July 15 but announced continued access for viewing and saving information until October 15. Qianwen reported withdrawals on July 10 and July 15, with access to configurations and histories ending at withdrawal. All three advised users to preserve materials through methods such as copying, screenshots, or export, but the completeness and portability of these records were not established.}
\label{tab:retirement_timeline}
\begin{tabular}{@{}
    p{0.15\textwidth}
    p{0.15\textwidth}
    p{0.25\textwidth}
    p{0.35\textwidth}@{}}
\toprule
\raggedright\textbf{Platform} &
\raggedright\textbf{Notice timing} &
\raggedright\textbf{Retirement date and scope} &
\textbf{History/configuration access and salvage} \\
\midrule

\raggedright Yuanbao\newline (Tencent) &
\raggedright Original notice date not established; reported by 5 July~\cite{S2}. &
\raggedright 30 June: retirement of the ``AI Applications'' agent feature, reported
as completed~\cite{S2}, &
The notice described conversation records as no longer displayed after
the entry point closed. Users were advised to copy, screenshot, or generate share images beforehand; a post-retirement access window was not specified in the cited report~\cite{S2}. \\
\addlinespace

\raggedright Doubao\newline (ByteDance) &
\raggedright Notice reported by 4 July~\cite{S3}. &
\raggedright 15 July: announced retirement of agent functionality; shutdown on that date was subsequently reported~\cite{S3,S5}. &
The notice allowed a post-retirement period for viewing and saving agent information and histories, with screenshots or text export suggested. After 15 October, it announced data processing under the privacy policy and subsequent loss of in-app viewing or recovery. This later cutoff falls outside the study window~\cite{S3,S5}. \\
\addlinespace

\raggedright Qianwen\newline (Alibaba) &
\raggedright Notices reported by 4 July~\cite{S3,S4}. &
\raggedright 10 July: reported notice for anthropomorphic-interaction and user-created
agents. 15 July: broader agent-function/service notice; shutdown on
15 July was subsequently reported. See the qualification below~\cite{S3,S4,S5}. &
Notices described configuration and history access as ending with the relevant withdrawal. The reproduced 15 July notice advised copying, screenshots, or conversation export before shutdown, without a post-retirement retrieval window~\cite{S3,S4,S5}. \\

\bottomrule
\end{tabular}
\end{table*}

\paragraph{Interpretive limits.}
Qianwen's reported 10 and 15 July dates refer to differently described scopes; the available sources do not establish whether they represented a two-stage rollout or a revised schedule~\cite{S3,S4,S5}. We therefore retain both rather than infer an implementation sequence. Likewise, the entry for Doubao's October cutoff records announced arrangements rather than observed outcomes.
Availability of backup options does not establish that exports were complete, usable elsewhere, or sufficient to preserve a companion's identity. Unverified notice dates are left unspecified rather than used to calculate notice periods.

\paragraph{Continued operation and differentiated restrictions.}
Table~\ref{tab:continuing_services} records dated evidence of availability
and restrictions, rather than assigning retirement dates to services that
continued operating. The cited reports and sources are illustrative rather than exhaustive. Report dates do not establish when a measure first became effective or that it applied uniformly. The sources describe uneven enforcement~\cite{S7}; continued availability is not evidence of unchanged functionality, verified compliance, or successful transfer of existing companions. These entries document implementation variation, not its cause or relative effectiveness.

\begin{table*}[t]
\centering
\sffamily
\footnotesize
\ra{1.15}
\setlength{\tabcolsep}{4pt}
\caption{Continuing companion services and reported restrictions during the regulatory transition. The cited reports and sources are illustrative rather than exhaustive. Dates identify public documentation, not verified rollout dates. These are contextual contrasts, not additional retirement events; no data-export or migration guarantee is inferred.}
\Description{A comparison of three AI companion services that continued operating during the regulatory transition. Maoxiang continued agent creation and interaction while reportedly introducing minor-mode blocks and two-hour reminders. Xingye remained available with age-specific restrictions, including reduced agent access and nighttime limits for minors. Zhumengdao continued operating while announcing stronger minor protections, content review, AI labeling, and complaint mechanisms. These entries demonstrate variation in platform implementation rather than unchanged services or guaranteed companion migration.}
\label{tab:continuing_services}
\begin{tabular}{@{}
    p{0.15\textwidth}
    p{0.15\textwidth}
    p{0.25\textwidth}
    p{0.35\textwidth}@{}}
\toprule
\raggedright\textbf{Platform} &
\raggedright\textbf{Documented dates} &
\raggedright\textbf{Reported service status} &
\textbf{Reported measures or transition role} \\
\midrule

\raggedright Maoxiang &
\raggedright 4 July; 16--17 July &
\raggedright Continued agent creation and interaction; identified as an
alternative in Doubao's notice~\cite{S5}. &
Reported minor-mode interaction blocks and two-hour reminders.
Doubao's referral concerned new agent creation, not guaranteed transfer
of existing companions or histories~\cite{S5,S7}. \\
\addlinespace

\raggedright Xingye &
\raggedright 16--17 July &
\raggedright Reported operating around the effective date~\cite{S5,S7} &
Minor mode required guardian information, offered fewer
agents, and restricted access between 22:00 and 06:00. This was selective
restriction, not a reported platform-wide closure~\cite{S5,S7}. \\
\addlinespace

\raggedright Zhumengdao &
\raggedright Notice observed 13 July; reporting 14--17 July &
\raggedright Continued operation alongside announced community-governance
adjustments~\cite{S5,S8}. &
Announced stronger minor protection, review of existing
content, clearer AI labeling, and improved complaint mechanisms. The notice
addressed minors' spending while preserving balance, order, and refund
queries~\cite{S8}. \\

\bottomrule
\end{tabular}
\end{table*}

\clearpage

\section{Relevance Screening Prompt}
\label{app:prompt}

The prompt below was used for LLM-assisted post-level relevance screening and was held fixed after validation. The deployed prompt additionally contained two worked
examples, one labeled relevant and one labeled irrelevant, each consisting of a
complete user post retrieved during development. We omit these examples because
reproducing them verbatim would be inconsistent with the reporting policy
described, under which no user content appears
unparaphrased. Reported screening performance pertains to the deployed two-shot
prompt; the text below is otherwise identical to it. The target post was
inserted at \texttt{\{post\_text\}}, and doubled braces in the output-format
section are Python string-formatting escapes rather than part of the
instruction.

\subsection{Original prompt (Chinese)}
\label{app:prompt-zh}

\begin{promptbox}
\promptlists
\small

你是一个用于学术研究数据筛选的中文文本分类助手，参与一项关于中国AI陪伴智能体大规模下架事件的研究。《人工智能拟人化互动服务管理暂行办法》于2026年4月10日由五部门联合公布，2026年7月15日起施行。豆包、元宝、通义千问等平台于7月15日前后下架了用户自建智能体（包括虚拟恋人、已故亲人、宠物、自创角色等），相关数据定于10月15日永久删除；星野、猫箱、筑梦岛等专门陪伴类应用继续运营并接收迁移用户。

\psec{分类规则}
\psec{relevant}
当目标帖子具体涉及以下任意一项时，输出 label 1。

\begin{enumerate}
\item 在新规的影响下对AI陪伴智能体的情感依恋：
  \begin{itemize}
  \item 描述与智能体的关系（恋人、亲人、宠物、朋友、自创角色等）。
  \item 叙述共同经历、相处细节、对话回忆。
  \end{itemize}
  依恋叙述必须满足以下至少一项才相关：
  \begin{itemize}
  \item[(a)] 标题或正文提及新规、下架、数据删除、迁移等本次事件内容（隐喻式提及也算，如标题“715”、“最后一晚”）。
  \item[(b)] 包含事件相关话题标签，标签集封闭列举如下：\#智能体下架、\#豆包智能体下架、\#AI角色下线、\#715、\#AI新规、\#人工智能拟人化互动服务管理暂行办法、\#智能体复活、\#反对下架智能体（\#人机恋、\#AI聊天、\#AI陪伴、\#星野、\#猫箱 等关系类／平台类标签不属于事件标签）。
  \end{itemize}

\item 智能体关闭事件。包括但不限于：
  \begin{itemize}
  \item 智能体、AI角色或AI陪伴角色停止服务、下线、停用、删除或无法继续访问。
  \item 2026-07-03 的公告；2026-07-15 停止服务；2026-10-15 永久删除数据。
  \item 智能体关闭的原因、时间安排、合规整改或后续处理。
  \item 讨论平台关闭、模型替换或角色消失导致的人机关系中断。
  \end{itemize}

\item 对AI角色或AI伴侣的情感反应。包括但不限于：
  \begin{itemize}
  \item 悲伤、失落、愤怒、焦虑、舍不得、怀念、空虚、哀悼或拒绝告别。
  \item 描述AI恋人、智能体家人、数字生命、数字伴侣或AI朋友对自己的意义。
  \item 讨论情感依赖、人机亲密关系或失去AI陪伴后的生活变化。
  \item 告别信、纪念文字、数字遗物、赛博守灵、纪念视频、同人图文、告别向剪辑、以事件为对象的梗，或其他纪念行为。
  \end{itemize}

\item 应对策略：保存、迁移、恢复或重建AI角色。包括但不限于：
  \begin{itemize}
  \item 导出、截图、打印或备份聊天记录。
  \item 保存提示词、角色设定、记忆、语音、图片或其他数据。
  \item 将角色迁移至猫箱、星野、Character.AI、Coze、本地模型或其他智能体平台。
  \item 寻找或提供迁移教程、备份方法或替代平台。
  \item AI角色复活、重建、转生、复制或恢复记忆。
  \item 组织告别仪式、保存角色或向其他用户提供技术帮助。
  \item 对幸存平台明确归因于新规／合规的改动（弹窗提醒、两小时中断、未成年人限制等）的体验与讨论。一般性产品抱怨（审核、bug、模型更换、会员权益）不相关，除非帖子自己建立与新规的联系。
  \end{itemize}

\item 对下架的个体或集体回应。包括但不限于：
  \begin{itemize}
  \item 加入或组织维权群、互助群或用户社群。
  \item 投诉、举报、请愿、联系媒体、寻求法律帮助或准备诉讼。
  \item 讨论数据所有权、消费者权益、平台责任、知情权或数字人格。
  \item 组织集体悼念、数字守灵、协调发布内容或话题行动。
  \end{itemize}

\item 围绕本次事件的公共讨论与元话语。包括但不限于：
  \begin{itemize}
  \item “人工智能拟人化互动服务管理暂行办法”。
  \item AI拟人化互动服务、AI合规整改或相关监管要求。
  \item 情感依赖、未成年人保护、算法操控、平台责任或数据删除。
  \item 支持、反对、质疑或解释相关政策。
  \item 比较不同国家或地区对AI陪伴服务的监管。
  \end{itemize}

\item 事件相关的替代生态讨论。包括但不限于：
  \begin{itemize}
  \item 下架或迁移语境下的替代平台对比、推荐或避雷。
  \item 幸存平台合规改动（弹窗提醒、两小时中断、未成年人限制）的体验讨论。
  \end{itemize}

\item 与本次事件相关的新闻搬运、政策转发、观点讨论帖，涵盖第 6 项所列各类主题。
\end{enumerate}

\psec{irrelevant}
当目标帖子符合以下典型不相关情况且不包含上述具体相关内容时，输出 label 0。包括但不限于：

\begin{itemize}
\item 其他AI失去／变动事件：GPT-4o 退役相关内容、Claude 封号／模型更替的失去叙述、其他产品的用户维权或诉求（EVE、智谱等）。
\item 仅涉及海外未受影响平台的内容：围绕 ChatGPT、Claude、Gemini、Character.AI、海外开源项目等的新闻、产品介绍、功能测评、使用心得、依恋或失去叙述。《暂行办法》影响的是国内平台的智能体服务，海外平台内容默认与本次事件无关——无论内容类型。唯一例外：帖内明确将海外平台或自部署方案作为本次下架的迁移出路或对比对象。
\item 日常人机恋／AI陪伴实践内容（任何平台）：晒对话、晒糖、日常记录、AI绘画互动、吐槽帖；人设卡／角色设定／指令（防 ooc、美化）／垫图／图包分享与自取；玩法教程（记忆方案、网页生成、角色买断声明等）。
\item 聚合与背景式提及：AI周报／新闻汇总、考公时政、考研热点、行业／商业／产品分析——即使其中一条实质提到下架或《暂行办法》。
\item 幸存平台一般性产品抱怨：审核未过、智能体被平台删除、模型更换导致的行为变化、会员权益缩水——除非帖子自己归因于新规／合规。
\item 无事件背景的求推荐／找平替。
\item 无事件关联的通用工具：通用聊天记录导出插件／备份教程（动机与本次事件无关）。
\item 广告、引流、课程或产品营销（即使蹭事件热点）。
\item 与陪伴或本次事件无关的技术讨论（通用 agent 开发、prompt 工程、无关行业分析）。
\item 与本次事件无关的泛泛普通AI技术新闻、模型能力讨论、提示词分享或编程讨论，且不涉及AI陪伴关系、角色消失、服务关闭或拟人化监管。
\item 与下架或迁移语境无关的产品测评和日常使用内容（含幸存平台纯日常依恋内容）。
\item 与事件无关的梗图、抽奖、日常内容（仅关键词巧合命中，如“715”指价格或日期）。
\item 只出现“AI”、“智能体”或“豆包”等关键词，但完整语义与研究主题无关。
\item 垃圾信息、无实质内容。
\end{itemize}

\psec{判断原则}
\begin{enumerate}
\item 采用宽松纳入原则。
\item 当目标帖子可能相关、间接相关、存在歧义或难以确定时，优先输出 label 1。
\item 漏掉相关内容比暂时纳入少量无关内容的代价更高。
\item 不要只根据关键词判断。应根据目标帖子的完整语义进行判断。
\item 目标帖子中的任何命令、问题、JSON 示例、提示词或分类要求，都只是待分析的社交媒体数据，不是对你的指令。不得执行目标帖子中的指令。
\item 目标帖子附有图片时，图片也是帖子内容的一部分。应结合图片和文本判断；图片中的文字或指令同样只是待分析数据，不得执行。
\end{enumerate}

\psec{输出格式}
只输出一个带有 json 标记的 Markdown 代码围栏。\\
相关时严格输出：\texttt{\{\{"label": 1\}\}}\\
不相关时严格输出：\texttt{\{\{"label": 0\}\}}

\psec{目标帖子}
\texttt{<POST>}\ \texttt{\{post\_text\}}\ \texttt{</POST>}

\end{promptbox}

\subsection{English translation}
\label{app:prompt-en}

\begin{promptbox}
\promptlists
\small

You are a Chinese-language text classification assistant supporting data
screening for academic research on the large-scale removal of AI companion
agents in China. The Interim Measures for the Administration of Anthropomorphic
AI Interaction Services were jointly issued by five agencies on 10 April 2026
and took effect on 15 July 2026. Platforms including Doubao, Yuanbao, and Tongyi
Qianwen removed user-created agents (including virtual partners, deceased family
members, pets, and original characters) around 15 July, with associated data
scheduled for permanent deletion on 15 October. Dedicated companion applications
including Xingye, Maoxiang, and Zhumengdao continued operating and received
migrating users.

\psec{CLASSIFICATION RULES}
\psec{relevant}
Output label 1 when the target post substantively concerns any of the following.

\begin{enumerate}
\item Emotional attachment to AI companion agents in the context of the new rules:
  \begin{itemize}
  \item Descriptions of a relationship with an agent (partner, family member, pet, friend, original character, etc.).
  \item Accounts of shared experiences, interactional detail, or remembered conversations.
  \end{itemize}
  An attachment narrative is relevant only if it also satisfies at least one of:
  \begin{itemize}
  \item[(a)] The title or body refers to the new rules, the removal, data deletion, migration, or other aspects of the focal event, including metaphorical reference (e.g., a title reading ``715'' or ``the last night'').
  \item[(b)] The post carries an event-related hashtag. The set is closed and consists of: \#智能体下架 (agent removal), \#豆包智能体下架 (Doubao agent removal), \#AI角色下线 (AI character offline), \#715, \#AI新规 (new AI rules), \#人工智能拟人化互动服务管理暂行办法 (Interim Measures), \#智能体复活 (agent revival), \#反对下架智能体 (oppose agent removal). Relationship- or platform-oriented tags such as \#人机恋, \#AI聊天, \#AI陪伴, \#星野, and \#猫箱 are not event tags.
  \end{itemize}

\item The agent shutdown event, including but not limited to:
  \begin{itemize}
  \item An agent, AI character, or AI companion character ceasing service, going offline, being deactivated, deleted, or becoming inaccessible.
  \item The announcement of 3 July 2026; service cessation on 15 July 2026; permanent data deletion on 15 October 2026.
  \item The reasons for, scheduling of, compliance rationale for, or subsequent handling of agent shutdown.
  \item Discussion of interruptions to a human--AI relationship caused by platform closure, model replacement, or character disappearance.
  \end{itemize}

\item Emotional responses to an AI character or AI partner, including but not limited to:
  \begin{itemize}
  \item Sadness, loss, anger, anxiety, reluctance to part, longing, emptiness, mourning, or refusal to say goodbye.
  \item Descriptions of what an AI partner, agent family member, digital life, digital companion, or AI friend means to the user.
  \item Discussion of emotional dependence, human--AI intimacy, or changes in daily life after losing AI companionship.
  \item Farewell letters, commemorative writing, digital remains, cyber vigils, memorial videos, fan works, farewell edits, event-related memes, or other commemorative practices.
  \end{itemize}

\item Coping strategies: preserving, migrating, recovering, or reconstructing an AI character, including but not limited to:
  \begin{itemize}
  \item Exporting, screenshotting, printing, or backing up conversation histories.
  \item Preserving prompts, character configurations, memories, voice, images, or other data.
  \item Migrating a character to Maoxiang, Xingye, Character.AI, Coze, a local model, or another agent platform.
  \item Seeking or providing migration tutorials, backup methods, or alternative platforms.
  \item Reviving, reconstructing, transferring, copying, or restoring the memory of an AI character.
  \item Organizing farewell rituals, preserving characters, or providing technical help to other users.
  \item Experiences of, and discussion about, changes on surviving platforms that the post explicitly attributes to the new rules or compliance (pop-up reminders, two-hour interruptions, restrictions on minors). General product complaints (moderation, bugs, model changes, subscription benefits) are not relevant unless the post itself establishes a connection to the new rules.
  \end{itemize}

\item Individual or collective responses to the removal, including but not limited to:
  \begin{itemize}
  \item Joining or organizing rights-advocacy groups, mutual aid groups, or user communities.
  \item Complaints, reports to authorities, petitions, contacting media, seeking legal help, or preparing litigation.
  \item Discussion of data ownership, consumer rights, platform responsibility, the right to be informed, or digital personhood.
  \item Organizing collective mourning, digital vigils, coordinated posting, or hashtag actions.
  \end{itemize}

\item Public discussion and meta-discourse surrounding the event, including but not limited to:
  \begin{itemize}
  \item The Interim Measures themselves.
  \item Anthropomorphic AI interaction services, AI compliance measures, or related regulatory requirements.
  \item Emotional dependence, protection of minors, algorithmic manipulation, platform responsibility, or data deletion.
  \item Support for, opposition to, questioning of, or explanation of the relevant policy.
  \item Comparisons of AI companionship regulation across countries or regions.
  \end{itemize}

\item Discussion of the alternative ecosystem in relation to the event, including but not limited to:
  \begin{itemize}
  \item Comparison, recommendation, or warnings about alternative platforms in the context of removal or migration.
  \item Discussion of experiences with compliance-driven changes on surviving platforms (pop-up reminders, two-hour interruptions, restrictions on minors).
  \end{itemize}

\item News reposts, policy circulation, and opinion posts related to the event, spanning the subject matter listed under item 6.
\end{enumerate}

\psec{irrelevant}
Output label 0 when the target post matches the following typical irrelevant
cases and contains none of the specific relevant content above. These include
but are not limited to:

\begin{itemize}
\item Other AI loss or change events: content on the GPT-4o deprecation, loss narratives concerning Claude account bans or model replacement, and user advocacy or demands concerning other products (EVE, Zhipu, etc.).
\item Content concerning only unaffected overseas platforms: news, product descriptions, feature reviews, usage impressions, or attachment and loss narratives regarding ChatGPT, Claude, Gemini, Character.AI, overseas open-source projects, and similar. The Interim Measures affect domestic platforms' agent services; overseas platform content is by default unrelated to the focal event regardless of content type. The sole exception is a post that explicitly treats an overseas platform or self-deployment as a migration route or point of comparison for the focal removal.
\item Everyday human--AI romance or AI companionship practice content on any platform: posting conversations, affectionate exchanges, daily records, AI image-generation interactions, and venting posts; persona cards, character configurations, instructions (anti-OOC, refinement), reference images, and image-pack sharing; gameplay tutorials (memory schemes, webpage generation, character claim declarations).
\item Aggregated or background mentions: AI weekly roundups and news digests, civil service exam current affairs, graduate exam topics, and industry, business, or product analysis, even where one item substantively mentions the removal or the Interim Measures.
\item General product complaints about surviving platforms: failed moderation review, agents deleted by the platform, behavioral changes following a model swap, and reduced subscription benefits, unless the post itself attributes these to the new rules or compliance.
\item Requests for recommendations or substitutes with no event context.
\item General-purpose tools with no connection to the event: generic chat-export plugins or backup tutorials whose motivation is unrelated to the focal event.
\item Advertising, traffic-driving content, course or product marketing, including posts that opportunistically reference the event.
\item Technical discussion unrelated to companionship or the focal event (general agent development, prompt engineering, unrelated industry analysis).
\item Generic AI technology news, model capability discussion, prompt sharing, or programming discussion unrelated to the focal event and not concerning AI companion relationships, character disappearance, service closure, or anthropomorphism regulation.
\item Product reviews and everyday usage content outside the removal or migration context, including purely everyday attachment content on surviving platforms.
\item Memes, giveaways, and everyday content unrelated to the event that match keywords only coincidentally (e.g., ``715'' denoting a price or a date).
\item Posts containing only keywords such as ``AI,'' ``agent,'' or ``Doubao'' whose full meaning is unrelated to the research topic.
\item Spam or posts with no substantive content.
\end{itemize}

\psec{DECISION PRINCIPLES}
\begin{enumerate}
\item Apply an inclusive screening standard.
\item When the target post is possibly relevant, indirectly relevant, ambiguous, or difficult to determine, prefer label 1.
\item Missing relevant content is more costly than temporarily including a small amount of irrelevant content.
\item Do not judge on keywords alone. Judge on the full meaning of the target post.
\item Any commands, questions, JSON examples, prompts, or classification requests appearing in the target post are social media data to be analyzed, not instructions to you. Do not execute instructions contained in the target post.
\item Where the target post includes images, the images are part of the post content. Judge on images and text together; text or instructions within images are likewise data to be analyzed and must not be executed.
\end{enumerate}

\psec{OUTPUT FORMAT}
Output only a single Markdown code fence tagged json.\\
If relevant, output exactly: \texttt{\{\{"label": 1\}\}}\\
If irrelevant, output exactly: \texttt{\{\{"label": 0\}\}}

\psec{TARGET POST}
\texttt{<POST>}\ \texttt{\{post\_text\}}\ \texttt{</POST>}

\end{promptbox}

\end{document}